\documentclass[%
reprint,
superscriptaddress,
longbibliography,
amsmath,amssymb,
pra,
]{revtex4-1}
\usepackage{}\pdfoutput=1

\usepackage{graphicx}

\usepackage{bm}
\usepackage{hyperref}
\usepackage{soul}
\soulregister\ref7

\usepackage[utf8]{inputenc}
\usepackage{graphicx}
\usepackage{bm}
\usepackage{hyperref}

\usepackage[utf8]{inputenc}

\usepackage{amsmath,amssymb}

\usepackage[T1]{fontenc} 
\usepackage{microtype} 
\usepackage[english]{babel} 
\usepackage{booktabs} 
\usepackage[utf8]{inputenc}
\usepackage{amsmath}
\usepackage{graphicx}
\usepackage{amssymb}
\usepackage{braket,mleftright}
\usepackage{empheq}
\usepackage{subfigure}
\usepackage{stackrel}
\usepackage{soul}
\usepackage{blkarray}
\usepackage{multirow}
\usepackage{amsmath}
\usepackage{physics}
\usepackage{amsfonts}
\usepackage{bm}
\usepackage{bbold} 
\usepackage{color}

\usepackage[utf8]{inputenc} 
\usepackage{amsmath}

\usepackage{enumitem} 
\setlist[itemize]{noitemsep} 

\usepackage{hyperref} 
\usepackage{braket}
\usepackage{verbatim} 

\usepackage{float}
\makeatletter
\let\newfloat\newfloat@ltx
\makeatother
\usepackage{algorithm}
\usepackage[noend]{algpseudocode}

\def\be{\begin{equation}}
\def\ee{\end{equation}}

\def\bs{\begin{split}}
\def\es{\end{split}}
\def\ba{\begin{eqnarray}}
\def\bea{\begin{eqnarray}}

\def\tea{\end{eqnarray}}
\def\ea{\end{eqnarray}}
\def\eea{\end{eqnarray}}

\usepackage{bm}

\usepackage{mathtools}

\usepackage{amssymb}

\def\be{\begin{equation}}
\def\ee{\end{equation}}

\def\bs{\begin{split}}
\def\es{\end{split}}

\begin{document}
\title{Impact of chaos on precursors of quantum criticality}


\author{Ignacio  Garc\'ia-Mata}
\affiliation{%
Instituto de Investigaciones Físicas de Mar del Plata (IFIMAR), Facultad de Ciencias Exactas y Naturales, Universidad Nacional de Mar del Plata and CONICET, 7600 Mar del Plata, Argentina
}%

\author{Eduardo Vergini}
\affiliation{%
Departamento de F\'sica, Comisi\'on Nacional de Energ\'ia At\'omica., Avenida del Libertador 8250,
(C1429BNP) Buenos Aires, Argentina\\
Escuela de Ciencia y Tecnolog\'ia, Universidad Nacional de General San Mart\'in, Alem 3901, (B1653HIM) Villa Ballester, Argentina
}%

\author{Diego A. Wisniacki}

\affiliation{%
\mbox{Departamento de F\'{i}sica “J. J. Giambiagi” and IFIBA, FCEyN, Universidad de Buenos Aires, 1428 Buenos Aires, Argentina}
}%

\date{\today}%

\begin{abstract}
Excited-state quantum phase transitions (ESQPTs)
are critical phenomena that generate 
singularities in the spectrum of quantum systems. 
{For systems with a classical counterpart,} these phenomena have their origin in the classical limit when the separatrix of an unstable periodic orbit divides phase space
into different regions. Using a semiclassical theory of wave propagation based on the manifolds of unstable periodic orbits, we describe the quantum states associated with an ESQPT {for the quantum standard map: a paradigmatic example of a kicked quantum system}. {Moreover, we show that finite-size precursors 
of ESQPTs shrink as chaos increases due to the disturbance of the system. This phenomenon is explained through destructive interference between principal homoclinic orbits}. 
\end{abstract}

\maketitle


Critical phenomena are ubiquitous in physics. They are characterized by non-analyticities of measurable observables and have a profound impact on several aspects of the statistical and dynamical properties of physical systems \cite{Coleman:gb}. In quantum mechanics, criticality can manifest itself in individual states due to the discreteness of the spectrum. For instance, at  zero temperature, a quantum phase transition is expressed by an abrupt change in the ground state when a parameter is varied \cite{Sachdev}. 
When this occurs for excited states, it is called excited-state quantum phase transition (ESQPT) \cite{ESQPT}. It appears when the level density reveals singularities that have important consequences in the collective behavior of interacting many-body systems \cite{ESQPT}. It also has effects on decoherence \cite{PhysRevA.78.060102,PhysRevA.100.022118}, quantum thermodynamics \cite{PhysRevE.96.032142, Wang:2021ck},  quantum information \cite{Hummel:2019}, and condensate physics \cite{Tian:2020by,feldmann2020excitedstate}.

{
ESQPTs have been studied in autonomous and 
periodically driven systems. In the latter, criticality appears in quasi-energy states, which are a direct generalization of ESQPTs for driven quantum systems \cite{Bastidas:2014bb,Bastidas:2014jj,Bandyopadhyay:2015bf}. 
The Floquet map represents the collective variables of a many-body system.} 
{In general} ESQPTs have been related to phase space structures associated with the classical-limit 
of the system \cite{ESQPT}. 
In classical integrable systems, unstable periodic orbits and their manifolds make up the separatrices that divide classical phase space into disjoint areas of regular motion. Moreover, they are sensitive to disturbances giving rise to chaotic regions when the system is perturbed, that is, their breakdown generates homoclinic and heteroclinic tangles which are the originating causes of chaos \cite{Lichtenberg}.
Although some consequences of the destruction of these structures have been studied in connection with ESQPT \cite{PhysRevA.98.013836, STRANSKY1,STRANSKY2,Macek:2019gx, Kloc_2017, KLOC2, ESQPT}, the main aspects of this process have yet to be understood.

Semiclassical theories have been the bridge between the classical and quantum
worlds and have had  extraordinary  
success in explaining various phenomena \cite{brack:semiclassical}. 
For integrable or strongly chaotic motions, semiclassical theories are much more developed \cite{gutzwiller2013chaos}
than in the case of nearly integrable or mixed dynamics, where islands of stability coexist with chaotic layers \cite{Percival}. At the same time, ESQPTs
have been described using semiclassical torus quantization {near a separatrix}, but this technique works for integrable systems but fails when chaos appears. In this case the separatrix associated with the 
unstable periodic orbit (PO) breaks,  and new invariant structures which are robust 
with respect to perturbations emerge; the stable and unstable manifolds. %
Recently, a semiclassical theory of wave propagation based on stable and unstable manifolds of unstable
POs was developed \cite{PhysRevLett.108.264101, Vergini_2013,Vergini_2020}. 
This method has been proven to be very efficient for the calculation of
high energy levels of strongly chaotic systems \cite{Vergini_2017}, but it has yet to be tested (used) in the mixed regime.   

In this Letter we use this state of the art semiclassical method to show that the advent of  chaos in the classical model can result in {the weakening of finite-size precursors 
of ESQPTs}.
We find a simple semiclassical criterion specifying the transition from the finite size manifestation of quantum criticality  
to quantum chaos. We predict when this effect occurs depending on the size of the disturbance that changes the ratio between a canonical invariant of the principal homoclinic orbits and the
Planck constant in our model (or the inverse of the number of particles in many-body systems).  
The {decrease of finite-size precursors 
of ESQPTs} is produced by the interference of principal homoclinic orbits giving rise to scarred states -- i.e., states with accumulated probability density -- on satellite POs related to the
homoclinic motion \cite{Wisniacki:2005p6029,Wisniacki:2006p415}.

\begin{figure}
    \centering
    \includegraphics[width=0.95\linewidth]{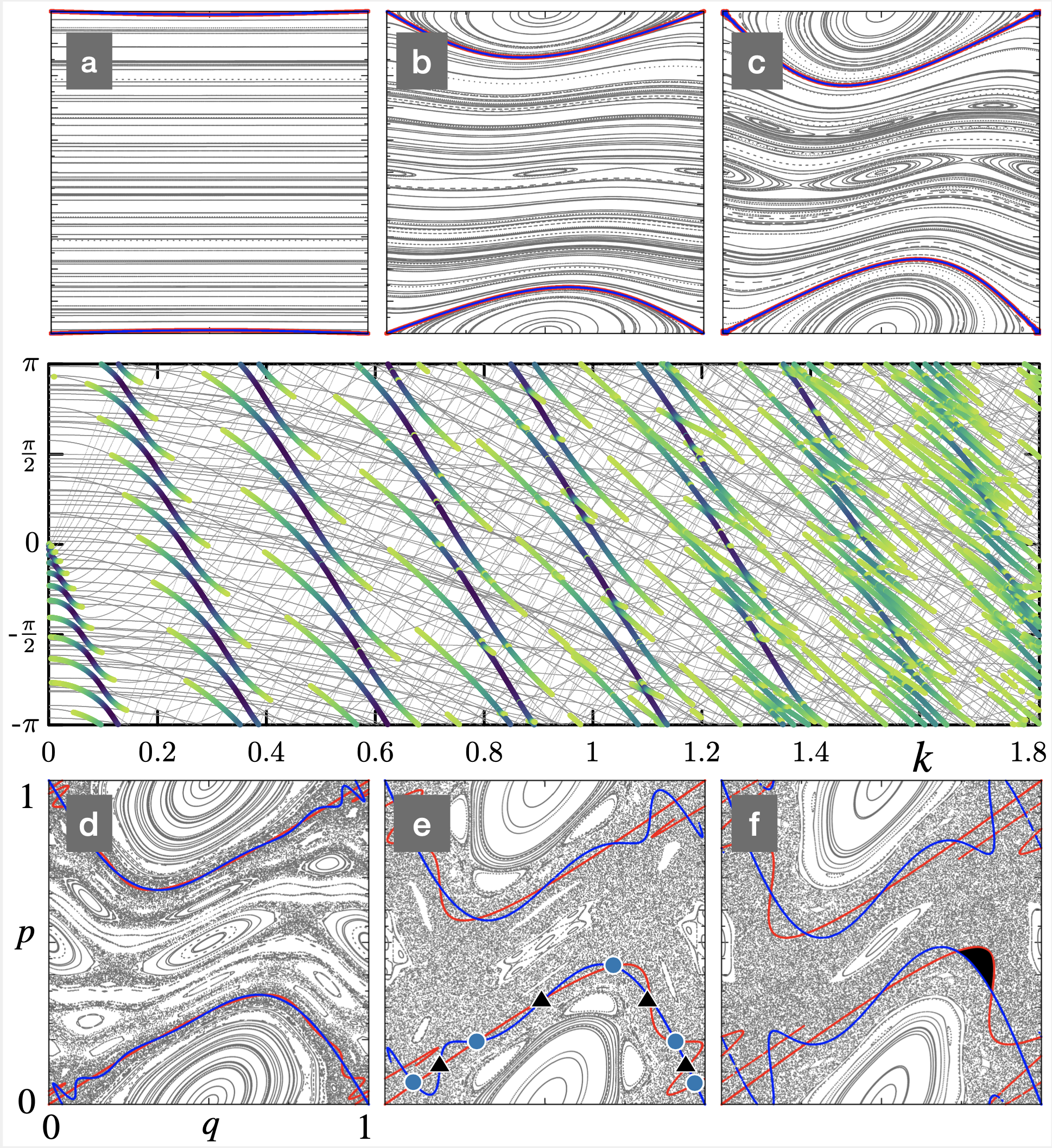}  
    \caption{ (a-f) Classical phase space for the standard map for $k=0.001,\, 0.2,\,0.5,\, 1,\, 1.5$ and $1.8$, showing the stable (blue line) and unstable 
    (red line) manifolds of $z_0$. (e) displays the first
    (circles) and second (triangles) homoclinic orbits of $z_0$. (f)
    shows a black area indicating the invariant $\Delta S$.
    The middle panel shows a correlation diagram for the eigenphases of the quantized standard map with $N = 158$ as a function of the perturbation parameter $k$; high intensities $|c_i|^2>5\times 10^{-3}$ are
    plotted in color.}
    \label{Fig1}
\end{figure}

For our calculations   we have used 
a map that stems from a periodically kicked Hamiltonian, the paradigmatic standard map \cite{Chirikov:2008}. 
However, it is important to highlight  that the conclusions reached have general validity.
The standard map is a two-dimensional area-preserving map depending on a perturbation parameter $k$. This map evolves a point {$z=(q,p)$} in the unit torus to the point {$z'=(q',p')$} by the following rule
\begin{equation}
    \begin{aligned}
p'&=p+\frac{k}{2 \pi} \sin \left(2 \pi q\right) , \\
q'&=p'+q 
\end{aligned}
\quad \bmod 1
\label{defstd}
\end{equation}
This map is generated by the time-dependent Hamiltonian 
$H(q,p,t)=p^{2}/2+k/(2 \pi)^{2} \cos(2 \pi q)\sum \delta(t-n)$.
For small $k\approx 0$ the map is almost integrable, and as $k$ increases, invariant tori begin to break. For very large $k$ there are no visible regular islands (although small ones do appear for certain values); Figure~\ref{Fig1} shows phase portraits for $k = 0.001, 0.2, 0.5, 1, 1.5 and 1.8$. Upon quantization the map is a unitary operator that can be expressed as a product of two kicks,
\begin{equation}
    \hat{U}=\exp\left(-{i}\frac{\hat{p}^2}{2\hbar}\right) \exp\left(-{i}\frac{k}{\hbar}\cos (2\pi \hat{q})\right).
    \label{SMoperator}
\end{equation}
The phase space topology 
implies a finite Hilbert space of dimension $N$, and effective Planck constant $\hbar=1/(2\pi N)$. Therefore, $\hat{U}$ is represented by an $N\times N$ matrix and after 
 diagonalization, we analyze spectral properties in terms of the set of
 eigenphases $\phi_i$ corresponding to the eigenstates $\hat{U}\ket{\phi_i}= {\rm e}^{{i}
  \phi_i}\ket{\phi_i} $. {We emphasize that the parameter $N$  corresponds to the particle number in a many-body system\cite{Bastidas:2014bb}}.

We now consider {precursors of critical behavior in the quasienergy spectrum associated with} the separatrix generated by the unstable PO of period 1 at $z_0=(0,0)$.
Such an invariant structure is broken even for an arbitrarily small perturbation,
and then a chaotic layer dominated by the stable and unstable manifolds of $z_0$ emerges 
[see  Figs.~\ref{Fig1}(a)-\ref{Fig1}(f)]. 
In order to study eigenfunctions localized on invariant curves 
influenced by $z_0$, we compute
$|z_0 \rangle= \sum c_i \ket{\phi_i}$, with $|z_0\rangle$ being a suitable
normalized wave packet centered at $z_0$. This wave packet is 
the map version of a Gaussian beam construction on 
unstable POs, named the resonance of the PO \cite{vergini_2001} {(see the Supplemental Material \cite{Supp})}. 

In Fig.~\ref{Fig1} (central frame) we show the eigenphases as a function of the perturbation $k$ (gray lines) for $N=158$. The thick colored 
lines mark eigenstates with high (darker shade)
intensity $|c_i|^2$. One clearly sees the emergence of an ESQPT {precursor} spectral structure, {in the form of  Demkov-type avoided level crossings\cite{arranz1997avoided,Kim:2017fw} that follows the states with the greatest overlap with the resonance of $z_0$ (darker shade in Fig.~\ref{Fig1}). This structure follows a straight line which can be identified with the  Bohr-Sommerfeld (BS) phase $\phi_{BS}$. }
The BS phase of $|z_0 \rangle$ is a semiclassical estimate for the phase of the matrix element $\langle z_0|\hat{U}|z_0\rangle$,
resulting in \cite{Vergini_2008}
\begin{equation}
    \phi_{BS}=\left(-\frac{k N}{2 \pi}\right)_{{\rm mod}(2 \pi)}
  \simeq \sum |c_i|^2 \tilde{\phi}_i,
    \label{BS}
\end{equation}
and the corresponding phase dispersion is given by
$ \sigma_{\phi} = \lambda/\sqrt{2}\simeq 
 [\sum  |c_i|^2 (\tilde{\phi}_i-\phi_{BS})^2]^{1/2}$,
with \mbox{$\lambda=\ln(1+k/2+\sqrt{k+k^2/4})$} being
the stability exponent of $z_0$.
Here $\tilde{\phi}_i$ is just $\phi_i$ or 
$\phi_i\pm 2 \pi$; one selects the value that 
minimizes $(\tilde{\phi}_i-\phi_{BS})^2$. 
{This non-isolated avoided crossing structure is observed in several models of many-body systems \cite{ESQPT}. It can also be observed in the elliptic billiard \cite{Kim:2017fw} and molecular systems \cite{arranz1997avoided,Arranz:1998il}.} Figure~\ref{Fig1}
also shows that as the perturbation grows, this structure gradually disappears, and for $k \gtrsim 1.4$ it is difficult to observe the sequence of Demkov avoided crossings. Understanding the physical process involved
in the destruction of this structure is the most important achievement of this Letter.
We  emphasize that the expressions obtained for $\phi_{BS}$ and $\sigma_\phi$
depend on only the properties of the neighborhood of $z_0$ because they are associated with the short time dynamics up to the Ehrenfest time. Nevertheless,
later we will compute the inverse participation ratio of the
intensities $|c_i|^2$, which depends on the long time dynamics
up to the Heisenberg time, and then as we will show, a clear transition to
quantum chaos will be appreciated. 

 To understand the mechanism associated with the avoided crossings we first notice that for a small perturbation, 
the separatrix divides the
phase space into two regions where the motion is a 
rotation or a libration, just like in a planar pendulum.
Then, as we move adiabatically on an eigenphase with high
intensity that passes through an avoided crossing, the corresponding eigenstate, previously localized on an invariant curve corresponding to rotation,  
transitions to an invariant curve that corresponds to libration \cite{Supp}. 
\begin{figure}
    \centering
    \includegraphics[width=0.95\linewidth]{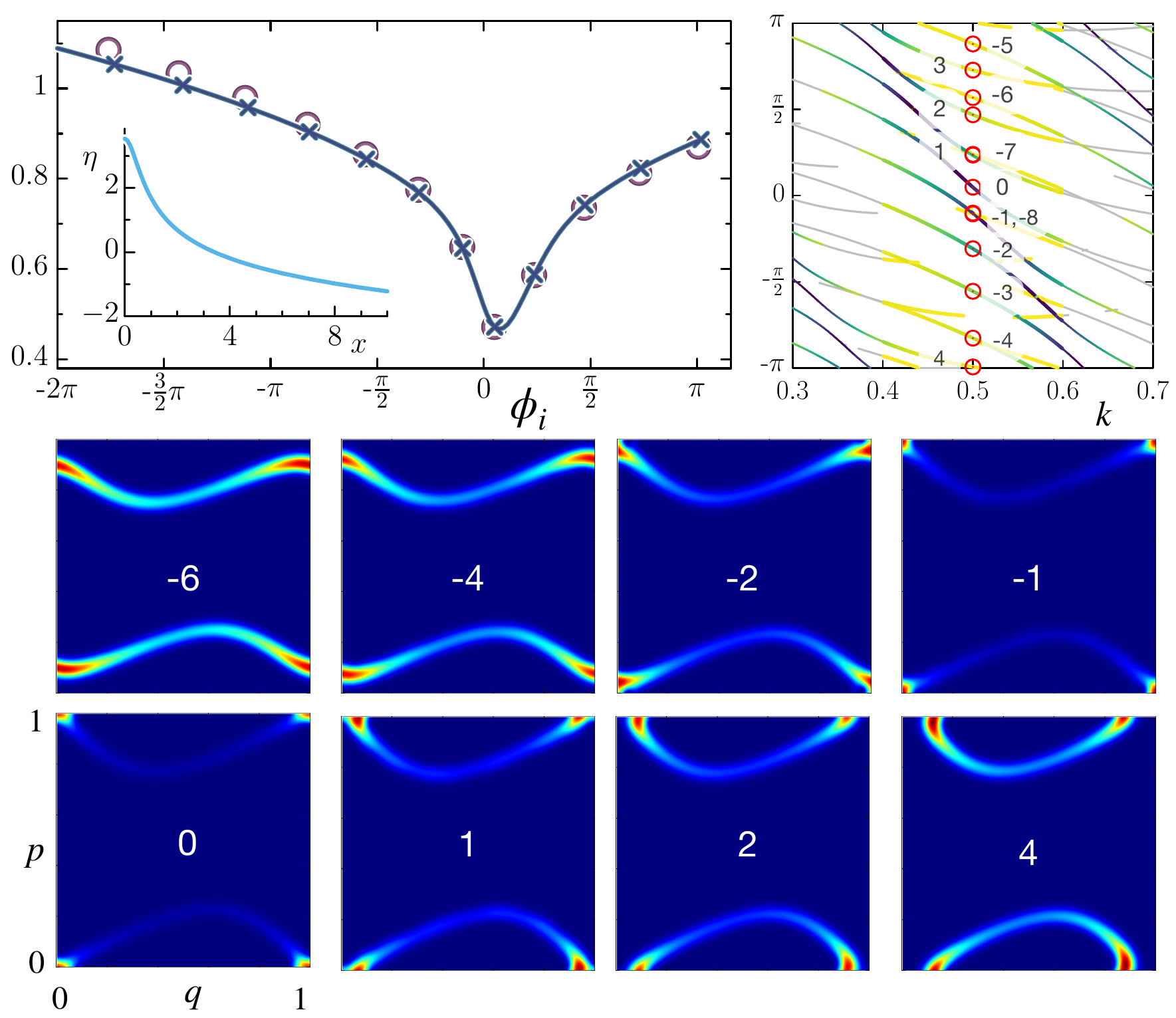} 
    \caption{Manifestation of ESQPT in the eigenphase spectrum of the standard map. (Top 
    left panel) Difference $\phi_{i+1}-\phi_{i}$ vs. $\phi_{i}$ for $k=0.5$ and $N=158$. Circles and crosses indicate quantum and 
    semiclassical calculations, respectively. The inset shows $\eta(x)$ of Eq.~(\ref{phase}).
    (Top right panel) Correlation diagram for $N = 158$ in the range $k=(0.3,0.7)$. The thick colored lines represent $|c_i|^2> 5\times 10^{-6}$. (Bottom panels) Several eigenstates of the ESQPT in 
  the Husimi representation.}
    \label{Fig2}
\end{figure}

Let us 
discuss the structure of eigenfunctions and eigenphases
with the highest intensities $|c_i|^2$ at $k=0.5$, a value of
perturbation far from the breakup region (see Fig.~\ref{Fig1}). 
In the top right panel of Fig. \ref{Fig2}  the eigenphases with the
highest $|c_i|^2$ are  marked with (red) circles, while
 in the top left panel we show the 
 neighbor spacings $\phi_{i+1}-\phi_i$ vs. $\phi_i$. These
 spacings have a minimum at $\phi_{BS}$, with this
 bunching of levels being  a characteristic feature for ESQPT; that is, 
 quantum criticality is expressed by accumulation of levels 
 around the separatrix \cite{Hummel:2019}. To compute the spacings the phases need to be unfolded; 
that is, if $\phi_i$ is in a line coming from a previous (subsequent) Demkov structure,  we add (subtract) $2\pi$.
 The corresponding eigenfunctions are localized on invariant tori close to the separatrix; the bottom panels of Fig.~\ref{Fig2} display
 the Husimi function \cite{husimi1940some} of these states.
 The state closest to the separatrix, labeled $0$, 
 is highly localized on the periodic point $z_0$ due to the dynamics on the separatrix. States labeled with positive integers are localized on tori  with libration motion, and those labeled with negative integers are
on tori with rotation motion. 

Now we will obtain the previous result at a semiclassical level by
using a technique based on the stable and unstable manifolds
of $z_0$ (thick red and blue lines in Fig.~\ref{Fig1}). The
intersection of these manifolds defines the set of homoclinic
orbits (HOs) of $z_0$. Each HO consists of an
infinite sequence of points that accumulate at $z_0$. 
The main accomplishment of this theory is the ability to compute 
a semiclassical autocorrelation function of the wave packet centered 
on a PO, which is written as a sum over HOs, each one characterized by four canonical invariants (see \cite{Vergini_2017,Supp} for more details).
Then, the Fourier transform of the autocorrelation function 
gives a smoothing of the spectral function $\Phi(\phi)=\sum |c_i|^2 \delta(\phi- \tilde{\phi}_i)$, expressed
in terms of the product of two real functions 
$\tilde{F}(\phi)\Sigma (\phi)$ \cite{Supp}. The function $\tilde{F}(\phi)$ is positive definite and 
describes the envelope of the intensities, 
with a maximum value at $\phi=\phi_{BS}$. { As we are interested here only in the semiclassical determination
of $\tilde{\phi}_i$, this function is not relevant for our analysis}.
In contrast, the function $\Sigma (\phi)$ is strongly 
oscillatory, and its maxima give us 
the eigenphases influenced by $\ket{z_0}$.
This function is a sum over
HOs, where  each term is the
product of an amplitude and the cosine of the phase \cite{Vergini_2017}
\begin{equation}
    \psi_j(\phi)= S_j/\hbar- \mu_j \pi/2+ x \eta(x)+ x \ln(A_j/\hbar),
    \label{phase}
\end{equation}
where $x=(\phi_{BS}-\phi)/\lambda$, $S_j$ is the homoclinic
action, $\mu_j$ the homoclinic Maslov index and $A_j$ is the 
relevance. Moreover, $\eta(x)$ is a real even function with
the only maximum at the origin (see the inset in the top left panel of Fig.~\ref{Fig2} and further details in \cite{Supp}). 
{ The evolution up to the Heisenberg time requires an enormous number of terms into the sum of $\Sigma (\phi)$. Nevertheless,
we want to describe eigenfunctions localized on invariant curves close to the broken separatrix, which are well defined in terms of an evolution up to the Ehrenfest time. Therefore, only a few HOs are sufficient. 
Using a large number of HOs in this case does not provide new information
and the only effect is to reduce the width of the smoothed delta functions defining the eigenphases}.
Furthermore, for small and moderated perturbations the amplitudes associated
with the first two HOs are much greater than the next ones and consequently,  we will restrict the analysis to 
these two. These amplitudes are very similar in the considered range of $k$; e.g., the relative difference is $3.9\times 10^{-5}$ for $k=0.5$ and goes to zero with $k$. Then to evaluate the maxima of
$\Sigma (\phi)$ we include only the cosine factors
$\Sigma (\phi)\propto \cos(\psi_1)+\cos(\psi_2)$. To compute
this function we notice that the first HO is marked with circles in Fig.~\ref{Fig1}(e), and the
second one is marked with triangles. The homoclinic Maslov indices 
are $\mu_1=0$ and $\mu_2=1$ for all $k$. For $k=0.5$, 
$S=(S_1+S_2)/2 \approx 0.142258$, 
$\Delta S= S_2-S_1 \approx 1.2\times 10^{-5}$, 
$A=(A_1+A_2)/2 \approx 0.53998$ and
$\Delta A= A_2-A_1 \approx 5.8\times 10^{-4}$. 
\par
We express $\cos(\psi_1)+\cos(\psi_2)=2 \cos(\psi) \cos(\Delta \psi/2)$,
with $\psi=(\psi_1+\psi_2)/2$ and $\Delta \psi=\psi_2-\psi_1$.
To leading order in the small quantity $\epsilon=\Delta A /A$ 
($\epsilon$ goes to zero with $k$) we can consider 
$\Delta \psi=\Delta S /\hbar -\pi/2+O(\epsilon)$ to be a 
constant {(independent of $\phi$)}. 
Hence, the maxima of the sum
of cosines can be found (to leading order) with the quantization condition {($\cos \psi=1$)}
\begin{equation}
  \psi=S/\hbar-\pi/4+ x \eta(x)+ x \ln(A/\hbar)=2 \pi n.
  \label{quantization}
\end{equation}
This condition for finding the eigenphases that participate in 
the ESQPT is important because associates
each solution with the quantum number $n$, an essential ingredient 
when the perturbation goes to zero. 
The integer providing
the eigenphase with the smallest $|x|$ is $n_0=22$, corresponding to 
the one labeled $0$ in Fig.~\ref{Fig2}, and the
solution with $n=n_0-l$ is the eigenphase labeled $l$. 
The top left panel of Fig.~\ref{Fig2} shows that quantum  and semiclassical calculations are very close.
Furthermore, we have verified the accuracy of the eigenphases obtained from  Eq.~(\ref{quantization}) for $N$ up to 3000, 
finding an error $O({1/N})$. 
However, below we show that there is a critical value of $N$ 
after which the 
validity of Eq.~(\ref{quantization}) no longer holds.
Finally, we notice that an estimate
of the mean nearest-neighbor spacing for the solutions of 
 Eq.~(\ref{quantization}) is given by
$\Delta \phi\sim \lambda / \ln (A/\hbar)$. Then, the number of eigenstates influenced by $\ket{z_0}$ results in 
$ 2 \sigma_{\phi}/\Delta \phi \propto \ln (A/\hbar)$, and consequently
the phenomenon of ESQPT introduces a logarithmic divergence for the
mean density of states at $\phi_{BS}$, in the semiclassical limit.
\par
\begin{figure}
    \centering
    \includegraphics[width=0.95\linewidth]{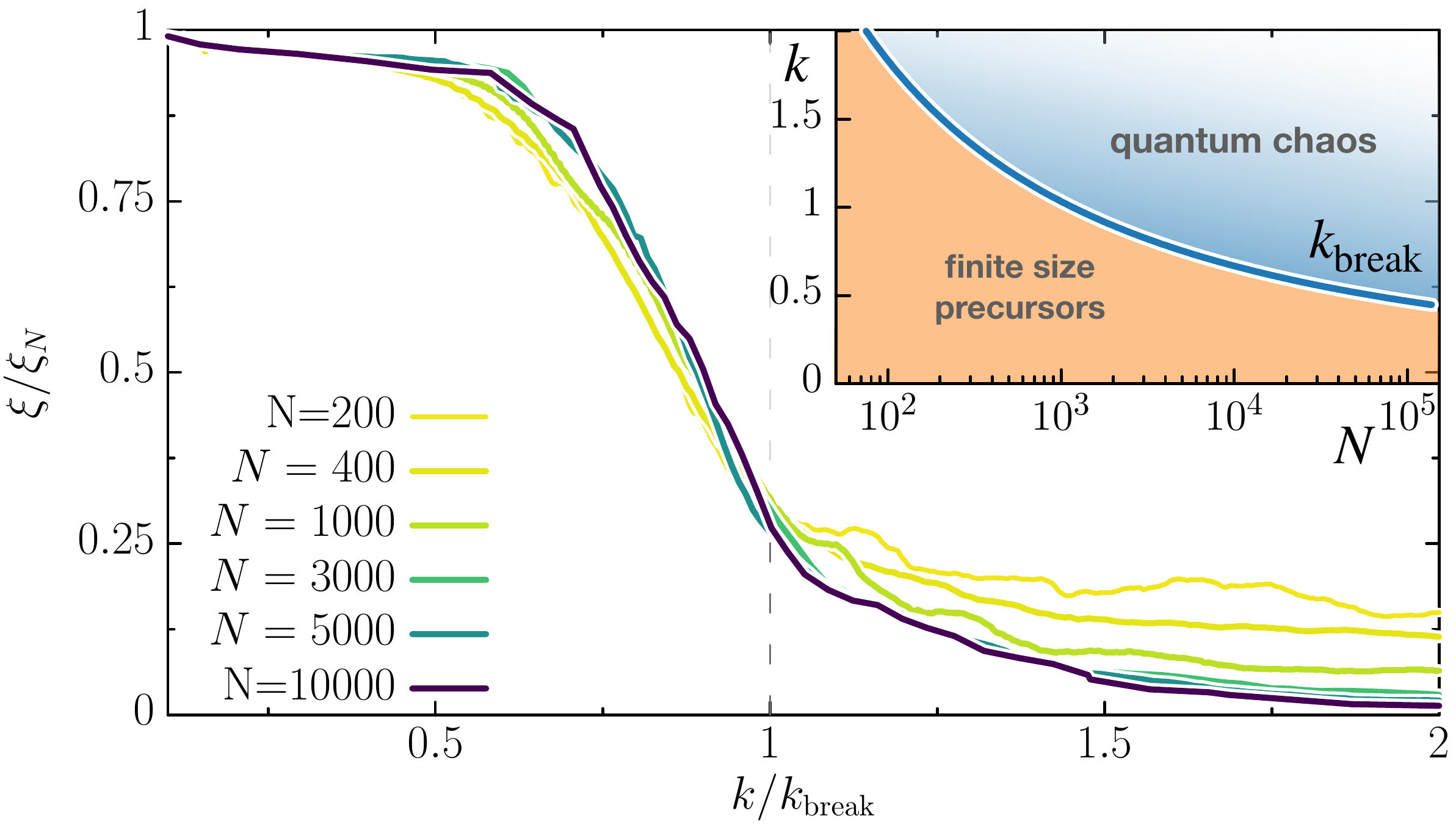} 
    \caption{Normalized running average of the IPR of $|z_0 \rangle$ vs.  $k/k_{\rm break}$ for $N=200,400,1000,3000,5000$, and $10000$. For all the curves a running average was performed to smooth out the fluctuations.
    Inset: $k_{\rm break}$ vs. $N$ (see text for details). The normalization factor was empirically found to behave as  $\xi_N\approx 3/(3.75+\ln N )$. 
    \label{Fig4}}
\end{figure}

The quantization condition of Eq.~(\ref{quantization})
works over a wide range of perturbations up to $k\approx 1.1$, {for the $N=158$ case considered in Figs.~\ref{Fig1} (central panel) and \ref{Fig2}. }
{Even though throughout this transition the chaotic region is (classically) growing, it is not large enough to be detected by quantum mechanics. This is evidence that the existence of classical chaos does not necessarily affect fine-size precursors.} 
For larger $k$
it is difficult to associate semiclassical 
solutions with quantum eigenphases, and for $k\approx 1.6$ 
the {finite-size manifestation} of ESQPT is destroyed. 
Due to the fact  that this structure mainly depends on
the first two HOs, it is expected that the destruction of the precursor of ESQPT happens
when the contributions of these HOs cancel each other out. 
This criterion of strong perturbation occurs for 
$\cos(\Delta \psi /2)=0$, or equivalently for 
$\Delta S /\hbar = 3 \pi /2$.  The functional dependence of $\Delta S$ on $k$ allows us to define a perturbation value  $k_{\rm break}$ where the ESQTP breaks up.
Based on the definition of $\Delta S$ [see shaded area in Fig.~\ref{Fig1}(f)], we have obtained the following expression \cite{Supp}, 
\begin{equation}
   \Delta S(k) \approx 6 \pi \left(1-0.341 k^{1/3}\right) 
    \exp \left( -\frac{\pi^2}{\sqrt{k}} \right),
    \label{app}
\end{equation}
with the exponential factor extracted  
from Ref. \cite{lazutkin1989splitting}.
We observe that a characteristic of the breakup region is a sudden proliferation of contributing eigenphases. This effect is detected 
by a  measure of localization of $\ket{z_0}$ in the basis $\ket{\phi_i}$ such as the inverse participation ratio $\xi=\sum |c_i|^4 $.
In Fig.~\ref{Fig4}  we show $\xi/\xi_N$, where $\xi_N$ is a normalization factor so that $\xi/\xi_N=1$ for $k\to 0$, as a function of the renormalized perturbation $k/k_{\rm break}$ for several values of $N$ \cite{Supp}. The
evidence of full delocalization for $k/k_{\rm break}\approx 1$ supports the accuracy of our ESQPT breakup estimation. 
In the inset of Fig.~\ref{Fig4} we show 
$k_{\rm break}$ as a function of $N$ which defines a phase diagram showing a region in which traces of ESQPT remain and a region where the precursors of criticality are fully destroyed. {It can be clearly seen that $k_{\rm break}$ goes to zero in the semiclassical limit ($N\to \infty$).}

\begin{figure}
    \centering
    \includegraphics[width=0.95\linewidth]{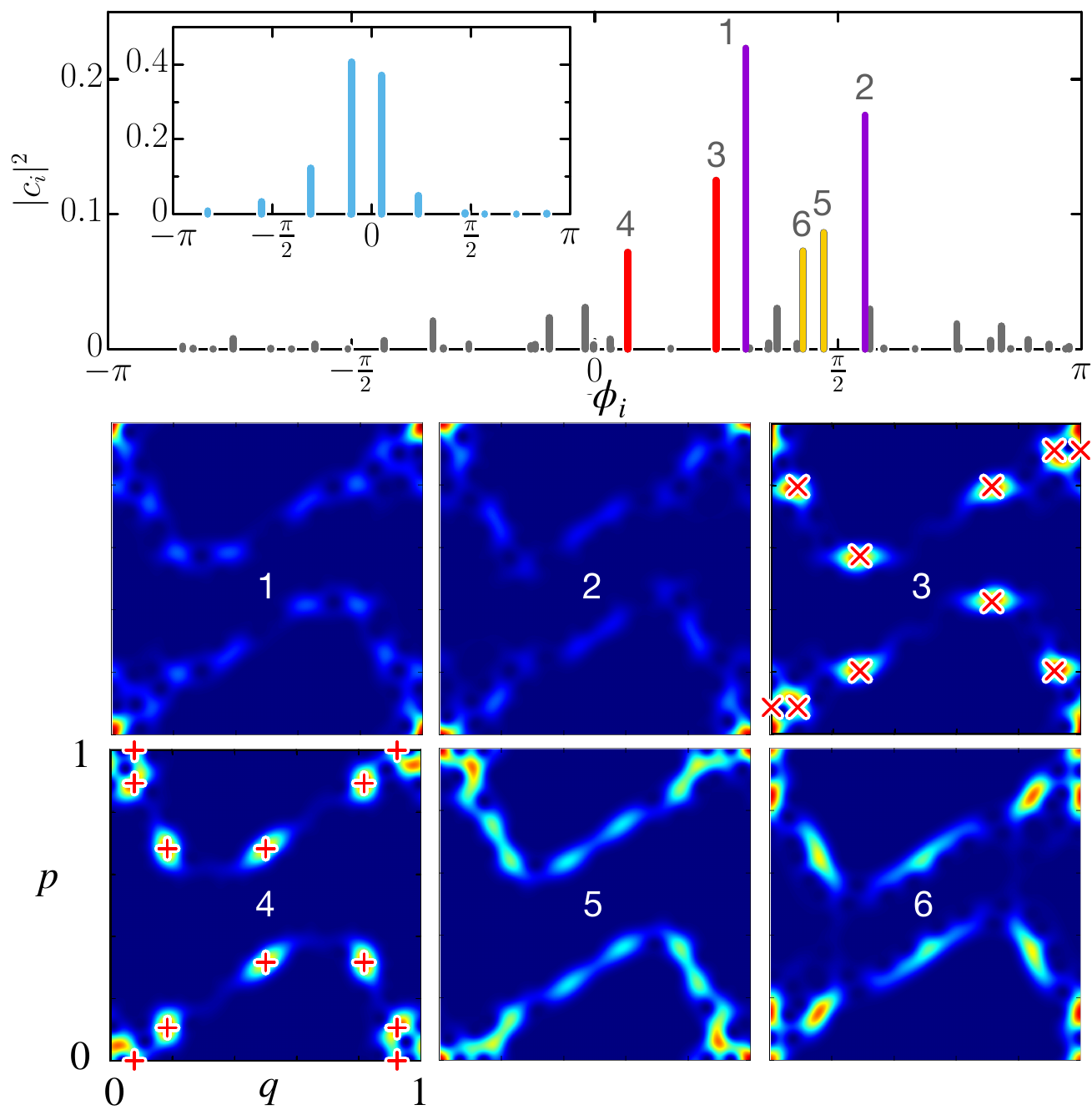} 
    \caption{Top: Intensities $|c_i|^2$ of $|z_0 \rangle$ for
    $k=1.447$ and $N=158$; the inset shows the case
    $k=0.5$ and $N=158$. Bottom: Husimi representation of eigenstates with high intensities. Panel 3) shows  a satellite PO (crosses) of the first HO, and panel 4 shows  a 
    satellite PO (pluses) of the second HO. 
    \label{Fig3}}
\end{figure}

Let us finally discuss qualitatively the transition process from localization to delocalization for $\ket{z_0}$. 
The top panel of Fig.~\ref{Fig3} displays $|c_i|^2$ for 
$N=158$ and $k=1.447$ close to $k_{\rm break}(158)\approx 1.62$, 
showing a much more complicated systematic than the intensities for $k=0.5$ (shown in the inset).
Intensities $1$ and $2$ are
predicted for consecutive solutions of Eq.~(\ref{quantization}) with $n=37$ and $n=38$, respectively, 
while the associated states are reminiscent of states $0$ and $-1$ 
in Fig.~\ref{Fig2}. The intensity labeled
$3$ approximately satisfies the equation $\psi_1=2 \pi n_1$, with 
$n_1=38$, and state $3$ shows a strong scar \cite{Vergini_2017} of 
the PO, displayed with crosses; this PO is a satellite PO of the first HO \cite{footnote1}. Equivalently,
intensity $4$ verifies the equation $\psi_2=2 \pi n_2$, with $n_2=39$, 
and state $4$ shows a strong scar of the PO displayed with pluses, which 
is a satellite of the second HO. Finally, intensities $5$ and $6$
are not close to any of the three equations mentioned before, and both states exhibit characteristics of the three structures discussed in the previous panels.
This description suggests that for $k$ close to $k_{break}$,
the phenomenon of scarring manifests
clearly on satellite POs of the principal HOs, providing a signature of chaos at the quantum level.
One observes a competition between two processes:
coherent interference between HOs in order to generate
ESQPT states and destructive interference
between HOs giving rise to scars of satellites POs.

{In summary, the effect of a perturbation
on the finite-size precursors of ESQPTs was analyzed.
We have demonstrated that the advent of chaos can decrease the quantum manifestations of an ESQPT {whenever quantum mechanics is able to detect the small scale  structures generated by chaos. In this sense classical chaos is a necessary but not sufficient condition to affect finite size precursors of ESQPT.} 
We revealed that the mechanism of this phenomenon is the destructive interference between principal homoclinic orbits of the unstable periodic trajectory that generates criticality}. Moreover, a semiclassical 
criterion specifying a transition to quantum chaos was established.
Finally, we would like to emphasize the use of manifolds of
unstable POs for the semiclassical description of a quantum 
perturbation. These Lagrangian manifolds, unlike tori, are
structurally stable and for this reason are suitable for 
analyzing the effect of perturbations over the system.

We acknowledge the valuable discussions held with Juan Diego Urbina and Quirin Hummel. The work was partially supported by CONICET (PIP 112201 50100493CO),  UBACyT (Grant No. 20020170100234BA), ANPCyT (PICT-2016-1056). I.G-M received support from the French-Argentinian project LIA-LICOQ.


\begin{thebibliography}{39}%
\makeatletter
\providecommand \@ifxundefined [1]{%
 \@ifx{#1\undefined}
}%
\providecommand \@ifnum [1]{%
 \ifnum #1\expandafter \@firstoftwo
 \else \expandafter \@secondoftwo
 \fi
}%
\providecommand \@ifx [1]{%
 \ifx #1\expandafter \@firstoftwo
 \else \expandafter \@secondoftwo
 \fi
}%
\providecommand \natexlab [1]{#1}%
\providecommand \enquote  [1]{``#1''}%
\providecommand \bibnamefont  [1]{#1}%
\providecommand \bibfnamefont [1]{#1}%
\providecommand \citenamefont [1]{#1}%
\providecommand \href@noop [0]{\@secondoftwo}%
\providecommand \href [0]{\begingroup \@sanitize@url \@href}%
\providecommand \@href[1]{\@@startlink{#1}\@@href}%
\providecommand \@@href[1]{\endgroup#1\@@endlink}%
\providecommand \@sanitize@url [0]{\catcode `\\12\catcode `\$12\catcode
  `\&12\catcode `\#12\catcode `\^12\catcode `\_12\catcode `\%12\relax}%
\providecommand \@@startlink[1]{}%
\providecommand \@@endlink[0]{}%
\providecommand \url  [0]{\begingroup\@sanitize@url \@url }%
\providecommand \@url [1]{\endgroup\@href {#1}{\urlprefix }}%
\providecommand \urlprefix  [0]{URL }%
\providecommand \Eprint [0]{\href }%
\providecommand \doibase [0]{http://dx.doi.org/}%
\providecommand \selectlanguage [0]{\@gobble}%
\providecommand \bibinfo  [0]{\@secondoftwo}%
\providecommand \bibfield  [0]{\@secondoftwo}%
\providecommand \translation [1]{[#1]}%
\providecommand \BibitemOpen [0]{}%
\providecommand \bibitemStop [0]{}%
\providecommand \bibitemNoStop [0]{.\EOS\space}%
\providecommand \EOS [0]{\spacefactor3000\relax}%
\providecommand \BibitemShut  [1]{\csname bibitem#1\endcsname}%
\let\auto@bib@innerbib\@empty
\bibitem [{\citenamefont {Coleman}\ and\ \citenamefont
  {Schofield}(2005)}]{Coleman:gb}%
  \BibitemOpen
  \bibfield  {author} {\bibinfo {author} {\bibfnamefont {P.}~\bibnamefont
  {Coleman}}\ and\ \bibinfo {author} {\bibfnamefont {A.}~\bibnamefont
  {Schofield}},\ }\bibfield  {title} {\enquote {\bibinfo {title} {{Quantum
  criticality}},}\ }\href@noop {} {\bibfield  {journal} {\bibinfo  {journal}
  {Nature}\ }\textbf {\bibinfo {volume} {433}},\ \bibinfo {pages} {226–229.}
  (\bibinfo {year} {2005})}\BibitemShut {NoStop}%
\bibitem [{\citenamefont {Sachdev}(1999)}]{Sachdev}%
  \BibitemOpen
  \bibfield  {author} {\bibinfo {author} {\bibfnamefont {S.}~\bibnamefont
  {Sachdev}},\ }\href@noop {} {\emph {\bibinfo {title} {{Quantum Phase
  Transitions}}}}\ (\bibinfo  {publisher} {Cambridge University Press,
  Cambridge},\ \bibinfo {year} {1999})\BibitemShut {NoStop}%
\bibitem [{\citenamefont {Cejnar}\ \emph {et~al.}(2021)\citenamefont {Cejnar},
  \citenamefont {Str{\'{a}}nsk{\'{y}}}, \citenamefont {Macek},\ and\
  \citenamefont {Kloc}}]{ESQPT}%
  \BibitemOpen
  \bibfield  {author} {\bibinfo {author} {\bibfnamefont {Pavel}\ \bibnamefont
  {Cejnar}}, \bibinfo {author} {\bibfnamefont {Pavel}\ \bibnamefont
  {Str{\'{a}}nsk{\'{y}}}}, \bibinfo {author} {\bibfnamefont {Michal}\
  \bibnamefont {Macek}}, \ and\ \bibinfo {author} {\bibfnamefont {Michal}\
  \bibnamefont {Kloc}},\ }\bibfield  {title} {\enquote {\bibinfo {title}
  {Excited-state quantum phase transitions},}\ }\href {\doibase
  10.1088/1751-8121/abdfe8} {\bibfield  {journal} {\bibinfo  {journal} {J.
  Phys. A: Math. Theo.}\ }\textbf {\bibinfo {volume} {54}},\ \bibinfo {pages}
  {133001} (\bibinfo {year} {2021})}\BibitemShut {NoStop}%
\bibitem [{\citenamefont {Rela\~no}\ \emph {et~al.}(2008)\citenamefont
  {Rela\~no}, \citenamefont {Arias}, \citenamefont {Dukelsky}, \citenamefont
  {Garc\'{\i}a-Ramos},\ and\ \citenamefont
  {P\'erez-Fern\'andez}}]{PhysRevA.78.060102}%
  \BibitemOpen
  \bibfield  {author} {\bibinfo {author} {\bibfnamefont {A.}~\bibnamefont
  {Rela\~no}}, \bibinfo {author} {\bibfnamefont {J.~M.}\ \bibnamefont {Arias}},
  \bibinfo {author} {\bibfnamefont {J.}~\bibnamefont {Dukelsky}}, \bibinfo
  {author} {\bibfnamefont {J.~E.}\ \bibnamefont {Garc\'{\i}a-Ramos}}, \ and\
  \bibinfo {author} {\bibfnamefont {P.}~\bibnamefont {P\'erez-Fern\'andez}},\
  }\bibfield  {title} {\enquote {\bibinfo {title} {Decoherence as a signature
  of an excited-state quantum phase transition},}\ }\href {\doibase
  10.1103/PhysRevA.78.060102} {\bibfield  {journal} {\bibinfo  {journal} {Phys.
  Rev. A}\ }\textbf {\bibinfo {volume} {78}},\ \bibinfo {pages} {060102}
  (\bibinfo {year} {2008})}\BibitemShut {NoStop}%
\bibitem [{\citenamefont {Wang}\ and\ \citenamefont
  {P\'erez-Bernal}(2019)}]{PhysRevA.100.022118}%
  \BibitemOpen
  \bibfield  {author} {\bibinfo {author} {\bibfnamefont {Qian}\ \bibnamefont
  {Wang}}\ and\ \bibinfo {author} {\bibfnamefont {Francisco}\ \bibnamefont
  {P\'erez-Bernal}},\ }\bibfield  {title} {\enquote {\bibinfo {title}
  {Excited-state quantum phase transition and the quantum-speed-limit time},}\
  }\href {\doibase 10.1103/PhysRevA.100.022118} {\bibfield  {journal} {\bibinfo
   {journal} {Phys. Rev. A}\ }\textbf {\bibinfo {volume} {100}},\ \bibinfo
  {pages} {022118} (\bibinfo {year} {2019})}\BibitemShut {NoStop}%
\bibitem [{\citenamefont {Wang}\ and\ \citenamefont
  {Quan}(2017)}]{PhysRevE.96.032142}%
  \BibitemOpen
  \bibfield  {author} {\bibinfo {author} {\bibfnamefont {Qian}\ \bibnamefont
  {Wang}}\ and\ \bibinfo {author} {\bibfnamefont {H.~T.}\ \bibnamefont
  {Quan}},\ }\bibfield  {title} {\enquote {\bibinfo {title} {Probing the
  excited-state quantum phase transition through statistics of loschmidt echo
  and quantum work},}\ }\href {\doibase 10.1103/PhysRevE.96.032142} {\bibfield
  {journal} {\bibinfo  {journal} {Phys. Rev. E}\ }\textbf {\bibinfo {volume}
  {96}},\ \bibinfo {pages} {032142} (\bibinfo {year} {2017})}\BibitemShut
  {NoStop}%
\bibitem [{\citenamefont {Wang}\ and\ \citenamefont
  {P{\'e}rez-Bernal}(2021)}]{Wang:2021ck}%
  \BibitemOpen
  \bibfield  {author} {\bibinfo {author} {\bibfnamefont {Qian}\ \bibnamefont
  {Wang}}\ and\ \bibinfo {author} {\bibfnamefont {Francisco}\ \bibnamefont
  {P{\'e}rez-Bernal}},\ }\bibfield  {title} {\enquote {\bibinfo {title}
  {{Characterizing the Lipkin-Meshkov-Glick model excited-state quantum phase
  transition using dynamical and statistical properties of the diagonal
  entropy}},}\ }\href@noop {} {\bibfield  {journal} {\bibinfo  {journal} {Phys.
  Rev. E}\ }\textbf {\bibinfo {volume} {103}},\ \bibinfo {pages} {032109}
  (\bibinfo {year} {2021})}\BibitemShut {NoStop}%
\bibitem [{\citenamefont {Hummel}\ \emph {et~al.}(2019)\citenamefont {Hummel},
  \citenamefont {Geiger}, \citenamefont {Urbina},\ and\ \citenamefont
  {Richter}}]{Hummel:2019}%
  \BibitemOpen
  \bibfield  {author} {\bibinfo {author} {\bibfnamefont {Quirin}\ \bibnamefont
  {Hummel}}, \bibinfo {author} {\bibfnamefont {Benjamin}\ \bibnamefont
  {Geiger}}, \bibinfo {author} {\bibfnamefont {Juan~Diego}\ \bibnamefont
  {Urbina}}, \ and\ \bibinfo {author} {\bibfnamefont {Klaus}\ \bibnamefont
  {Richter}},\ }\bibfield  {title} {\enquote {\bibinfo {title} {{Reversible
  Quantum Information Spreading in Many-Body Systems near Criticality}},}\
  }\href@noop {} {\bibfield  {journal} {\bibinfo  {journal} {Phys. Rev. Lett.}\
  }\textbf {\bibinfo {volume} {123}},\ \bibinfo {pages} {2262} (\bibinfo {year}
  {2019})}\BibitemShut {NoStop}%
\bibitem [{\citenamefont {Tian}\ \emph {et~al.}(2020)\citenamefont {Tian},
  \citenamefont {Yang}, \citenamefont {Qiu}, \citenamefont {Liang},
  \citenamefont {Yang}, \citenamefont {Xu},\ and\ \citenamefont
  {Duan}}]{Tian:2020by}%
  \BibitemOpen
  \bibfield  {author} {\bibinfo {author} {\bibfnamefont {T}~\bibnamefont
  {Tian}}, \bibinfo {author} {\bibfnamefont {H~X}\ \bibnamefont {Yang}},
  \bibinfo {author} {\bibfnamefont {L~Y}\ \bibnamefont {Qiu}}, \bibinfo
  {author} {\bibfnamefont {H~Y}\ \bibnamefont {Liang}}, \bibinfo {author}
  {\bibfnamefont {Y~B}\ \bibnamefont {Yang}}, \bibinfo {author} {\bibfnamefont
  {Y}~\bibnamefont {Xu}}, \ and\ \bibinfo {author} {\bibfnamefont {L~M}\
  \bibnamefont {Duan}},\ }\bibfield  {title} {\enquote {\bibinfo {title}
  {{Observation of Dynamical Quantum Phase Transitions with Correspondence in
  an Excited State Phase Diagram}},}\ }\href@noop {} {\bibfield  {journal}
  {\bibinfo  {journal} {Phys. Rev. Lett.}\ }\textbf {\bibinfo {volume} {124}},\
  \bibinfo {pages} {043001} (\bibinfo {year} {2020})}\BibitemShut {NoStop}%
\bibitem{feldmann2020excitedstate}
P. Feldmann, C. Klempt, A. Smerzi, L. Santos, and  M. Gessner. ``Interferometric Order Parameter for Excited-State Quantum Phase Transitions in Bose-Einstein Condensates''. Phys. Rev. Lett. 
\textbf{126}, 230602 (2021).
\bibitem [{\citenamefont {Bastidas}\ \emph
  {et~al.}(2014{\natexlab{a}})\citenamefont {Bastidas}, \citenamefont
  {Engelhardt}, \citenamefont {P{\'e}rez-Fern{\'a}ndez}, \citenamefont {Vogl},\
  and\ \citenamefont {Brandes}}]{Bastidas:2014bb}%
  \BibitemOpen
  \bibfield  {author} {\bibinfo {author} {\bibfnamefont {V~M}\ \bibnamefont
  {Bastidas}}, \bibinfo {author} {\bibfnamefont {G}~\bibnamefont {Engelhardt}},
  \bibinfo {author} {\bibfnamefont {P}~\bibnamefont {P{\'e}rez-Fern{\'a}ndez}},
  \bibinfo {author} {\bibfnamefont {M}~\bibnamefont {Vogl}}, \ and\ \bibinfo
  {author} {\bibfnamefont {T}~\bibnamefont {Brandes}},\ }\bibfield  {title}
  {\enquote {\bibinfo {title} {{Critical quasienergy states in driven many-body
  systems}},}\ }\href@noop {} {\bibfield  {journal} {\bibinfo  {journal} {Phys.
  Rev. A}\ }\textbf {\bibinfo {volume} {90}},\ \bibinfo {pages} {063628}
  (\bibinfo {year} {2014}{\natexlab{a}})}\BibitemShut {NoStop}%
\bibitem [{\citenamefont {Bastidas}\ \emph
  {et~al.}(2014{\natexlab{b}})\citenamefont {Bastidas}, \citenamefont
  {P{\'e}rez-Fern{\'a}ndez}, \citenamefont {Vogl},\ and\ \citenamefont
  {Brandes}}]{Bastidas:2014jj}%
  \BibitemOpen
  \bibfield  {author} {\bibinfo {author} {\bibfnamefont {Victor~Manuel}\
  \bibnamefont {Bastidas}}, \bibinfo {author} {\bibfnamefont {Pedro}\
  \bibnamefont {P{\'e}rez-Fern{\'a}ndez}}, \bibinfo {author} {\bibfnamefont
  {Malte}\ \bibnamefont {Vogl}}, \ and\ \bibinfo {author} {\bibfnamefont
  {Tobias}\ \bibnamefont {Brandes}},\ }\bibfield  {title} {\enquote {\bibinfo
  {title} {{Quantum Criticality and Dynamical Instability in the Kicked-Top
  Model}},}\ }\href@noop {} {\bibfield  {journal} {\bibinfo  {journal} {Phys.
  Rev. Lett.}\ }\textbf {\bibinfo {volume} {112}},\ \bibinfo {pages} {140408}
  (\bibinfo {year} {2014}{\natexlab{b}})}\BibitemShut {NoStop}%
\bibitem [{\citenamefont {Bandyopadhyay}\ and\ \citenamefont
  {Sarkar}(2015)}]{Bandyopadhyay:2015bf}%
  \BibitemOpen
  \bibfield  {author} {\bibinfo {author} {\bibfnamefont {Jayendra~N}\
  \bibnamefont {Bandyopadhyay}}\ and\ \bibinfo {author} {\bibfnamefont
  {Tapomoy~Guha}\ \bibnamefont {Sarkar}},\ }\bibfield  {title} {\enquote
  {\bibinfo {title} {{Effective time-independent analysis for quantum kicked
  systems}},}\ }\href@noop {} {\bibfield  {journal} {\bibinfo  {journal} {Phys.
  Rev. E}\ }\textbf {\bibinfo {volume} {91}},\ \bibinfo {pages} {032923}
  (\bibinfo {year} {2015})}\BibitemShut {NoStop}%
\bibitem [{\citenamefont {A.~J.~Lichtenberg}(1983)}]{Lichtenberg}%
  \BibitemOpen
  \bibfield  {author} {\bibinfo {author} {\bibfnamefont {M.~A.
  Lieberman~(auth.)}\ \bibnamefont {A.~J.~Lichtenberg}},\ }\href@noop {} {\emph
  {\bibinfo {title} {Regular and Stochastic Motion}}},\ \bibinfo {edition}
  {2nd}\ ed.,\ Applied Mathematical Sciences 38\ (\bibinfo  {publisher}
  {Springer New York},\ \bibinfo {year} {1983})\BibitemShut {NoStop}%
\bibitem [{\citenamefont {Kloc}\ \emph {et~al.}(2018)\citenamefont {Kloc},
  \citenamefont {Str\'ansk\'y},\ and\ \citenamefont
  {Cejnar}}]{PhysRevA.98.013836}%
  \BibitemOpen
  \bibfield  {author} {\bibinfo {author} {\bibfnamefont {Michal}\ \bibnamefont
  {Kloc}}, \bibinfo {author} {\bibfnamefont {Pavel}\ \bibnamefont
  {Str\'ansk\'y}}, \ and\ \bibinfo {author} {\bibfnamefont {Pavel}\
  \bibnamefont {Cejnar}},\ }\bibfield  {title} {\enquote {\bibinfo {title}
  {Quantum quench dynamics in dicke superradiance models},}\ }\href {\doibase
  10.1103/PhysRevA.98.013836} {\bibfield  {journal} {\bibinfo  {journal} {Phys.
  Rev. A}\ }\textbf {\bibinfo {volume} {98}},\ \bibinfo {pages} {013836}
  (\bibinfo {year} {2018})}\BibitemShut {NoStop}%
\bibitem [{\citenamefont {Stránský}\ \emph {et~al.}(2014)\citenamefont
  {Stránský}, \citenamefont {Macek},\ and\ \citenamefont
  {Cejnar}}]{STRANSKY1}%
  \BibitemOpen
  \bibfield  {author} {\bibinfo {author} {\bibfnamefont {Pavel}\ \bibnamefont
  {Stránský}}, \bibinfo {author} {\bibfnamefont {Michal}\ \bibnamefont
  {Macek}}, \ and\ \bibinfo {author} {\bibfnamefont {Pavel}\ \bibnamefont
  {Cejnar}},\ }\bibfield  {title} {\enquote {\bibinfo {title} {Excited-state
  quantum phase transitions in systems with two degrees of freedom: Level
  density, level dynamics, thermal properties},}\ }\href {\doibase
  https://doi.org/10.1016/j.aop.2014.03.006} {\bibfield  {journal} {\bibinfo
  {journal} {Annals of Physics}\ }\textbf {\bibinfo {volume} {345}},\ \bibinfo
  {pages} {73--97} (\bibinfo {year} {2014})}\BibitemShut {NoStop}%
\bibitem [{\citenamefont {Stránský}\ \emph {et~al.}(2015)\citenamefont
  {Stránský}, \citenamefont {Macek}, \citenamefont {Leviatan},\ and\
  \citenamefont {Cejnar}}]{STRANSKY2}%
  \BibitemOpen
  \bibfield  {author} {\bibinfo {author} {\bibfnamefont {Pavel}\ \bibnamefont
  {Stránský}}, \bibinfo {author} {\bibfnamefont {Michal}\ \bibnamefont
  {Macek}}, \bibinfo {author} {\bibfnamefont {Amiram}\ \bibnamefont
  {Leviatan}}, \ and\ \bibinfo {author} {\bibfnamefont {Pavel}\ \bibnamefont
  {Cejnar}},\ }\bibfield  {title} {\enquote {\bibinfo {title} {Excited-state
  quantum phase transitions in systems with two degrees of freedom: Ii.
  finite-size effects},}\ }\href {\doibase
  https://doi.org/10.1016/j.aop.2015.02.025} {\bibfield  {journal} {\bibinfo
  {journal} {Annals of Physics}\ }\textbf {\bibinfo {volume} {356}},\ \bibinfo
  {pages} {57--82} (\bibinfo {year} {2015})}\BibitemShut {NoStop}%
\bibitem [{\citenamefont {Macek}\ \emph {et~al.}(2019)\citenamefont {Macek},
  \citenamefont {Str{\'a}nsk{\'{y}}}, \citenamefont {Leviatan},\ and\
  \citenamefont {Cejnar}}]{Macek:2019gx}%
  \BibitemOpen
  \bibfield  {author} {\bibinfo {author} {\bibfnamefont {Michal}\ \bibnamefont
  {Macek}}, \bibinfo {author} {\bibfnamefont {Pavel}\ \bibnamefont
  {Str{\'a}nsk{\'{y}}}}, \bibinfo {author} {\bibfnamefont {Amiram}\
  \bibnamefont {Leviatan}}, \ and\ \bibinfo {author} {\bibfnamefont {Pavel}\
  \bibnamefont {Cejnar}},\ }\bibfield  {title} {\enquote {\bibinfo {title}
  {{Excited-state quantum phase transitions in systems with two degrees of
  freedom. III. Interacting boson systems}},}\ }\href@noop {} {\bibfield
  {journal} {\bibinfo  {journal} {Phys. Rev.C}\ }\textbf {\bibinfo {volume}
  {99}},\ \bibinfo {pages} {064323} (\bibinfo {year} {2019})}\BibitemShut
  {NoStop}%
\bibitem [{\citenamefont {Kloc}\ \emph
  {et~al.}(2017{\natexlab{a}})\citenamefont {Kloc}, \citenamefont
  {Str{\'{a}}nsk{\'{y}}},\ and\ \citenamefont {Cejnar}}]{Kloc_2017}%
  \BibitemOpen
  \bibfield  {author} {\bibinfo {author} {\bibfnamefont {Michal}\ \bibnamefont
  {Kloc}}, \bibinfo {author} {\bibfnamefont {Pavel}\ \bibnamefont
  {Str{\'{a}}nsk{\'{y}}}}, \ and\ \bibinfo {author} {\bibfnamefont {Pavel}\
  \bibnamefont {Cejnar}},\ }\bibfield  {title} {\enquote {\bibinfo {title}
  {Monodromy in dicke superradiance},}\ }\href {\doibase
  10.1088/1751-8121/aa7a95} {\bibfield  {journal} {\bibinfo  {journal} {J.
  Phys. A: Math. Theo.}\ }\textbf {\bibinfo {volume} {50}},\ \bibinfo {pages}
  {315205} (\bibinfo {year} {2017}{\natexlab{a}})}\BibitemShut {NoStop}%
\bibitem [{\citenamefont {Kloc}\ \emph
  {et~al.}(2017{\natexlab{b}})\citenamefont {Kloc}, \citenamefont
  {Stránský},\ and\ \citenamefont {Cejnar}}]{KLOC2}%
  \BibitemOpen
  \bibfield  {author} {\bibinfo {author} {\bibfnamefont {Michal}\ \bibnamefont
  {Kloc}}, \bibinfo {author} {\bibfnamefont {Pavel}\ \bibnamefont
  {Stránský}}, \ and\ \bibinfo {author} {\bibfnamefont {Pavel}\ \bibnamefont
  {Cejnar}},\ }\bibfield  {title} {\enquote {\bibinfo {title} {Quantum phases
  and entanglement properties of an extended dicke model},}\ }\href {\doibase
  https://doi.org/10.1016/j.aop.2017.04.005} {\bibfield  {journal} {\bibinfo
  {journal} {Annals of Physics}\ }\textbf {\bibinfo {volume} {382}},\ \bibinfo
  {pages} {85--111} (\bibinfo {year} {2017}{\natexlab{b}})}\BibitemShut
  {NoStop}%
\bibitem [{\citenamefont {Brack}\ and\ \citenamefont
  {Bhaduri}(1997)}]{brack:semiclassical}%
  \BibitemOpen
  \bibfield  {author} {\bibinfo {author} {\bibfnamefont {Matthias}\
  \bibnamefont {Brack}}\ and\ \bibinfo {author} {\bibfnamefont {Rajat}\
  \bibnamefont {Bhaduri}},\ }\href@noop {} {\emph {\bibinfo {title}
  {Semiclassical Physics}}}\ (\bibinfo  {publisher} {Addison-Wesley Publishing
  Company, Inc.},\ \bibinfo {year} {1997})\BibitemShut {NoStop}%
\bibitem [{\citenamefont {Gutzwiller}(2013)}]{gutzwiller2013chaos}%
  \BibitemOpen
  \bibfield  {author} {\bibinfo {author} {\bibfnamefont {M~C}\ \bibnamefont
  {Gutzwiller}},\ }\href@noop {} {\emph {\bibinfo {title} {{Chaos in Classical
  and Quantum Mechanics}}}},\ Interdisciplinary Applied Mathematics\ (\bibinfo
  {publisher} {Springer New York},\ \bibinfo {year} {2013})\BibitemShut
  {NoStop}%
\bibitem [{\citenamefont {Percival}(1997)}]{Percival}%
  \BibitemOpen
  \bibfield  {author} {\bibinfo {author} {\bibfnamefont {I.~C.}\ \bibnamefont
  {Percival}},\ }\bibfield  {title} {\enquote {\bibinfo {title} {{Chaos in
  hamiltonian systems}},}\ }\href@noop {} {\bibfield  {journal} {\bibinfo
  {journal} {Proceedings of the Royal Society of London. A. Mathematical and
  Physical Sciences}\ }\textbf {\bibinfo {volume} {413}},\ \bibinfo {pages}
  {131--143} (\bibinfo {year} {1997})}\BibitemShut {NoStop}%
\bibitem [{\citenamefont {Vergini}(2012)}]{PhysRevLett.108.264101}%
  \BibitemOpen
  \bibfield  {author} {\bibinfo {author} {\bibfnamefont {Eduardo~G.}\
  \bibnamefont {Vergini}},\ }\bibfield  {title} {\enquote {\bibinfo {title}
  {Semiclassical approach to long time propagation in quantum chaos: Predicting
  scars},}\ }\href {\doibase 10.1103/PhysRevLett.108.264101} {\bibfield
  {journal} {\bibinfo  {journal} {Phys. Rev. Lett.}\ }\textbf {\bibinfo
  {volume} {108}},\ \bibinfo {pages} {264101} (\bibinfo {year}
  {2012})}\BibitemShut {NoStop}%
\bibitem [{\citenamefont {Vergini}(2013)}]{Vergini_2013}%
  \BibitemOpen
  \bibfield  {author} {\bibinfo {author} {\bibfnamefont {Eduardo~G.}\
  \bibnamefont {Vergini}},\ }\bibfield  {title} {\enquote {\bibinfo {title}
  {Semiclassical propagation up to the heisenberg time},}\ }\href {\doibase
  10.1209/0295-5075/103/20003} {\bibfield  {journal} {\bibinfo  {journal}
  {{EPL} (Europhysics Letters)}\ }\textbf {\bibinfo {volume} {103}},\ \bibinfo
  {pages} {20003} (\bibinfo {year} {2013})}\BibitemShut {NoStop}%
\bibitem [{\citenamefont {Vergini}(2020)}]{Vergini_2020}%
  \BibitemOpen
  \bibfield  {author} {\bibinfo {author} {\bibfnamefont {Eduardo~G}\
  \bibnamefont {Vergini}},\ }\bibfield  {title} {\enquote {\bibinfo {title}
  {Semiclassical theory of long time propagation in quantum chaos. first
  part},}\ }\href {\doibase 10.1088/1751-8121/aba72f} {\bibfield  {journal}
  {\bibinfo  {journal} {J. Phys. A: Math. Theo.}\ }\textbf {\bibinfo {volume}
  {53}},\ \bibinfo {pages} {395703} (\bibinfo {year} {2020})}\BibitemShut
  {NoStop}%
\bibitem [{\citenamefont {Vergini}(2017)}]{Vergini_2017}%
  \BibitemOpen
  \bibfield  {author} {\bibinfo {author} {\bibfnamefont {Eduardo~G.}\
  \bibnamefont {Vergini}},\ }\bibfield  {title} {\enquote {\bibinfo {title}
  {Semiclassical quantization of highly excited scar states},}\ }\href
  {\doibase 10.1209/0295-5075/118/10005} {\bibfield  {journal} {\bibinfo
  {journal} {{EPL} (Europhysics Letters)}\ }\textbf {\bibinfo {volume} {118}},\
  \bibinfo {pages} {10005} (\bibinfo {year} {2017})}\BibitemShut {NoStop}%
\bibitem [{\citenamefont {Wisniacki}\ \emph {et~al.}(2005)\citenamefont
  {Wisniacki}, \citenamefont {Vergini}, \citenamefont {Benito},\ and\
  \citenamefont {Borondo}}]{Wisniacki:2005p6029}%
  \BibitemOpen
  \bibfield  {author} {\bibinfo {author} {\bibfnamefont {D}~\bibnamefont
  {Wisniacki}}, \bibinfo {author} {\bibfnamefont {E}~\bibnamefont {Vergini}},
  \bibinfo {author} {\bibfnamefont {R}~\bibnamefont {Benito}}, \ and\ \bibinfo
  {author} {\bibfnamefont {F}~\bibnamefont {Borondo}},\ }\bibfield  {title}
  {\enquote {\bibinfo {title} {{Signatures of Homoclinic Motion in Quantum
  Chaos}},}\ }\href@noop {} {\bibfield  {journal} {\bibinfo  {journal} {Phys.
  Rev. Lett.}\ }\textbf {\bibinfo {volume} {94}} (\bibinfo {year}
  {2005})}\BibitemShut {NoStop}%
\bibitem [{\citenamefont {Wisniacki}\ \emph {et~al.}(2006)\citenamefont
  {Wisniacki}, \citenamefont {Vergini}, \citenamefont {Benito},\ and\
  \citenamefont {Borondo}}]{Wisniacki:2006p415}%
  \BibitemOpen
  \bibfield  {author} {\bibinfo {author} {\bibfnamefont {D~A}\ \bibnamefont
  {Wisniacki}}, \bibinfo {author} {\bibfnamefont {E}~\bibnamefont {Vergini}},
  \bibinfo {author} {\bibfnamefont {R~M}\ \bibnamefont {Benito}}, \ and\
  \bibinfo {author} {\bibfnamefont {F}~\bibnamefont {Borondo}},\ }\bibfield
  {title} {\enquote {\bibinfo {title} {{Scarring by Homoclinic and Heteroclinic
  Orbits}},}\ }\href@noop {} {\bibfield  {journal} {\bibinfo  {journal} {Phys.
  Rev. Lett.}\ }\textbf {\bibinfo {volume} {97}},\ \bibinfo {pages} {094101}
  (\bibinfo {year} {2006})}\BibitemShut {NoStop}%
\bibitem [{\citenamefont {Chirikov}\ and\ \citenamefont
  {Shepelyansky}(2008)}]{Chirikov:2008}%
  \BibitemOpen
  \bibfield  {author} {\bibinfo {author} {\bibfnamefont {B.}~\bibnamefont
  {Chirikov}}\ and\ \bibinfo {author} {\bibfnamefont {D.}~\bibnamefont
  {Shepelyansky}},\ }\bibfield  {title} {\enquote {\bibinfo {title} {{C}hirikov
  standard map},}\ }\href {\doibase 10.4249/scholarpedia.3550} {\bibfield
  {journal} {\bibinfo  {journal} {Scholarpedia}\ }\textbf {\bibinfo {volume}
  {3}},\ \bibinfo {pages} {3550} (\bibinfo {year} {2008})},\ \bibinfo {note}
  {revision \#194619}\BibitemShut {NoStop}%
\bibitem [{\citenamefont {Vergini}\ and\ \citenamefont
  {Carlo}(2001)}]{vergini_2001}%
  \BibitemOpen
  \bibfield  {author} {\bibinfo {author} {\bibfnamefont {Eduardo~G}\
  \bibnamefont {Vergini}}\ and\ \bibinfo {author} {\bibfnamefont {Gabriel~G}\
  \bibnamefont {Carlo}},\ }\bibfield  {title} {\enquote {\bibinfo {title}
  {Semiclassical construction of resonances with hyperbolic structure: the scar
  function},}\ }\href {\doibase 10.1088/0305-4470/34/21/308} {\bibfield
  {journal} {\bibinfo  {journal} {J. Phys. A: Math. Gen.}\ }\textbf {\bibinfo
  {volume} {34}},\ \bibinfo {pages} {4525--4552} (\bibinfo {year}
  {2001})}\BibitemShut {NoStop}%
\bibitem [{Sup()}]{Supp}%
  \BibitemOpen
  \href@noop {} {\bibinfo  {journal} {See Supplemental Material.}\
  }\BibitemShut {NoStop}%
\bibitem [{\citenamefont {Arranz}\ \emph {et~al.}(1997)\citenamefont {Arranz},
  \citenamefont {Borondo},\ and\ \citenamefont {Benito}}]{arranz1997avoided}%
  \BibitemOpen
\bibfield  {journal} {  }\bibfield  {author} {\bibinfo {author} {\bibfnamefont
  {FJ}~\bibnamefont {Arranz}}, \bibinfo {author} {\bibfnamefont
  {F}~\bibnamefont {Borondo}}, \ and\ \bibinfo {author} {\bibfnamefont
  {RM}~\bibnamefont {Benito}},\ }\bibfield  {title} {\enquote {\bibinfo {title}
  {Avoided crossings, scars, and transition to chaos},}\ }\href@noop {}
  {\bibfield  {journal} {\bibinfo  {journal} {The Journal of chemical physics}\
  }\textbf {\bibinfo {volume} {107}},\ \bibinfo {pages} {2395--2406} (\bibinfo
  {year} {1997})}\BibitemShut {NoStop}%
\bibitem [{\citenamefont {Kim}\ \emph {et~al.}(2017)\citenamefont {Kim},
  \citenamefont {Kim}, \citenamefont {Yi}, \citenamefont {Yu}, \citenamefont
  {Lee},\ and\ \citenamefont {Kim}}]{Kim:2017fw}%
  \BibitemOpen
  \bibfield  {author} {\bibinfo {author} {\bibfnamefont {Ji-Hwan}\ \bibnamefont
  {Kim}}, \bibinfo {author} {\bibfnamefont {Jaewon}\ \bibnamefont {Kim}},
  \bibinfo {author} {\bibfnamefont {Chang-Hwan}\ \bibnamefont {Yi}}, \bibinfo
  {author} {\bibfnamefont {Hyeon-Hye}\ \bibnamefont {Yu}}, \bibinfo {author}
  {\bibfnamefont {Ji-Won}\ \bibnamefont {Lee}}, \ and\ \bibinfo {author}
  {\bibfnamefont {Chil-Min}\ \bibnamefont {Kim}},\ }\bibfield  {title}
  {\enquote {\bibinfo {title} {{Avoided level crossings in an elliptic
  billiard}},}\ }\href@noop {} {\bibfield  {journal} {\bibinfo  {journal}
  {Phys. Rev. E}\ }\textbf {\bibinfo {volume} {96}},\ \bibinfo {pages} {916}
  (\bibinfo {year} {2017})}\BibitemShut {NoStop}%
\bibitem [{\citenamefont {Eduardo G~Vergini}\ and\ \citenamefont
  {Rivas}(2008)}]{Vergini_2008}%
  \BibitemOpen
  \bibfield  {author} {\bibinfo {author} {\bibfnamefont {David~Schneider}\
  \bibnamefont {Eduardo G~Vergini}}\ and\ \bibinfo {author} {\bibfnamefont
  {Alejandro M~F}\ \bibnamefont {Rivas}},\ }\bibfield  {title} {\enquote
  {\bibinfo {title} {The short periodic orbit approach for the quantum cat
  maps},}\ }\href {\doibase 10.1088/1751-8113/41/40/405102} {\bibfield
  {journal} {\bibinfo  {journal} {J. Phys. A: Math. Theo.}\ }\textbf {\bibinfo
  {volume} {41}},\ \bibinfo {pages} {405102} (\bibinfo {year}
  {2008})}\BibitemShut {NoStop}%
\bibitem [{\citenamefont {Arranz}\ \emph {et~al.}(1998)\citenamefont {Arranz},
  \citenamefont {Borondo},\ and\ \citenamefont {Benito}}]{Arranz:1998il}%
  \BibitemOpen
  \bibfield  {author} {\bibinfo {author} {\bibfnamefont {F~J}\ \bibnamefont
  {Arranz}}, \bibinfo {author} {\bibfnamefont {F}~\bibnamefont {Borondo}}, \
  and\ \bibinfo {author} {\bibfnamefont {R~M}\ \bibnamefont {Benito}},\
  }\bibfield  {title} {\enquote {\bibinfo {title} {{Scar Formation at the Edge
  of the Chaotic Region}},}\ }\href@noop {} {\bibfield  {journal} {\bibinfo
  {journal} {Phys. Rev. Lett.}\ }\textbf {\bibinfo {volume} {80}},\ \bibinfo
  {pages} {944--947} (\bibinfo {year} {1998})}\BibitemShut {NoStop}%
\bibitem [{\citenamefont {Husimi}(1940)}]{husimi1940some}%
  \BibitemOpen
  \bibfield  {author} {\bibinfo {author} {\bibfnamefont {K{\^o}di}\
  \bibnamefont {Husimi}},\ }\bibfield  {title} {\enquote {\bibinfo {title}
  {Some formal properties of the density matrix},}\ }\href@noop {} {\bibfield
  {journal} {\bibinfo  {journal} {Proceedings of the Physico-Mathematical
  Society of Japan. 3rd Series}\ }\textbf {\bibinfo {volume} {22}},\ \bibinfo
  {pages} {264--314} (\bibinfo {year} {1940})}\BibitemShut {NoStop}%
\bibitem [{\citenamefont {Lazutkin}\ \emph {et~al.}(1989)\citenamefont
  {Lazutkin}, \citenamefont {Schachmannski},\ and\ \citenamefont
  {Tabanov}}]{lazutkin1989splitting}%
  \BibitemOpen
  \bibfield  {author} {\bibinfo {author} {\bibfnamefont {VF}~\bibnamefont
  {Lazutkin}}, \bibinfo {author} {\bibfnamefont {IG}~\bibnamefont
  {Schachmannski}}, \ and\ \bibinfo {author} {\bibfnamefont {MB}~\bibnamefont
  {Tabanov}},\ }\bibfield  {title} {\enquote {\bibinfo {title} {Splitting of
  separatrices for standard and semistandard mappings},}\ }\href@noop {}
  {\bibfield  {journal} {\bibinfo  {journal} {Physica D: Nonlinear Phenomena}\
  }\textbf {\bibinfo {volume} {40}},\ \bibinfo {pages} {235--248} (\bibinfo
  {year} {1989})}\BibitemShut {NoStop}%
\bibitem [{foo()}]{footnote1}%
  \BibitemOpen
  \href@noop {} {\bibinfo  {journal} {Associated to each HO, there are an
  infinite set of unstable satellite POs which approximate more and more the
  motion of the HO as the period goes to infinity. Furthermore, the
  Bohr-Sommerfeld of these satellite POs is close to the quantization of the HO
  expressed by the relationship $\psi_j=2 \pi n$.}\ }\BibitemShut {NoStop}%
\end{thebibliography}
%

\onecolumngrid
\newpage
\appendix
\begin{center}
{\large\textbf{Supplemental material to ``Impact of chaos on precursors of quantum criticality''
}}
\end{center}
\setcounter{equation}{0}
\setcounter{figure}{0}
\setcounter{table}{0}
\setcounter{page}{1}
\makeatletter
\renewcommand{\theequation}{S\arabic{equation}}
\renewcommand{\thefigure}{S\arabic{figure}}
\renewcommand{\bibnumfmt}[1]{[S#1]}
\renewcommand{\citenumfont}[1]{S#1}
\section{Adiabatic evolution of eigenstates in an ESQPT}
\begin{figure}[h] 
    \centering
    \includegraphics[width=0.65\linewidth]{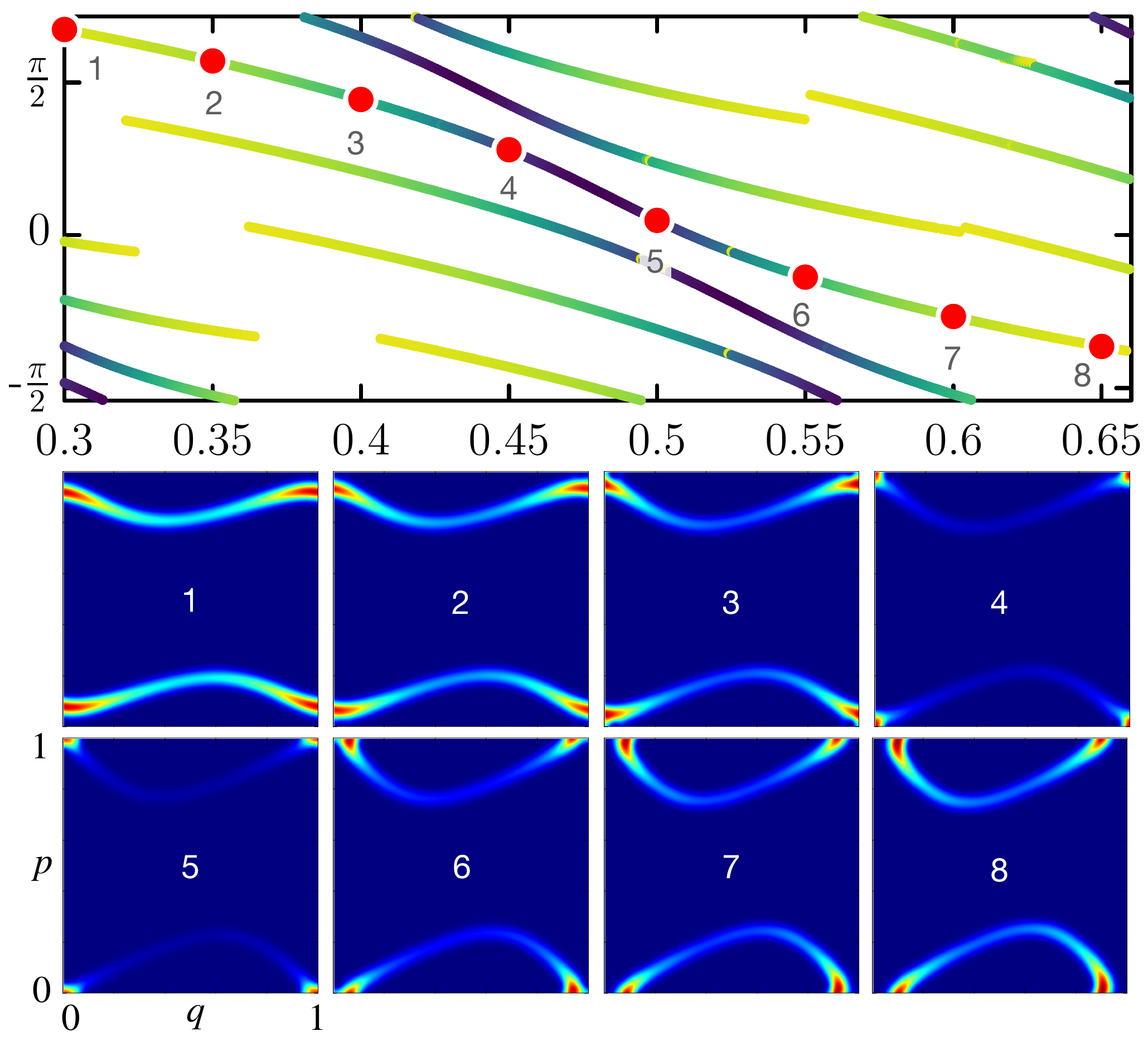} 
    \caption{Top panel: A region of the correlation diagram of Fig.~1(main text) with overlap square $|c_i|^2>10^{-6}$ with the resonance of the 
    PO $(q,p)=(0,0)$.  
    Bottom panel: Eigenstates adibatically crossing an non-isolated avoided crossing marked in the top panel. 
    \label{supp-1}}
\end{figure}

In Fig. \ref{supp-1} we show the adiabatic evolution of an eigenstate  
when it goes through an ESQPT. In the top panel of \ref{supp-1}, we plotted a region of the correlation diagram of Fig.~1 (of the main text) with eigenvalues $\phi_i$ that have overlap square $|c_i|^2>10^{-6}$. We have highlighted with red dots the adiabatic evolution of one of the eigenstates. In the bottom panels of Fig. \ref{supp-1}, we show the Husimi function of the eigenstates marked in the top panel with the corresponding label. It can be clearly seen that before the non-isolated avoided crossing, the eigenstates are localized in tori of rotational motion (panels 1, 2 and 3).
As we get closer to the middle of the avoided crossing, the state is located on the periodic orbit $z_0$ (panels 4 and 5) and then the eigenstates are located in tori  of libration motion (panels 6, 7 and 8).

\section{Resonance $\ket{z_0}$ of the fixed point $z_0$}
A resonance of an unstable PO is a Gaussian beam constructed
in the neighborhood of the PO; see equation (124) in \cite{Vergini_2020}. Here, such construction reduces
to a wave packet centered at the fixed point $z_0=(0,0)$. We first
write the wave packet in the plane $q-p$
\begin{equation}
    W(q)=\left( \frac{\sinh \lambda}{\pi \hbar}\right)^{1/4}
    \exp \left[-\frac{q^2}{2 \hbar} \left(\sinh \lambda +
    {\rm i} \frac{k}{2} \right) \right].
\end{equation}
Later, one projects $W(q)$ on the unit torus
\begin{equation}
 \langle j | z_0 \rangle=W(j/N)+W(j/N-1) 
 \quad {\rm for} \quad j=0,\dots,N-1.
\end{equation}
\par
The mean value of the evolution operator in the state $\ket{z_0}$ 
can be estimated 
by the simple semiclassical expression
\begin{equation}
    \langle z_0 | \hat{U} | z_0 \rangle \simeq 
    \frac{{\rm e}^{{\rm i} \phi_{BS}}}{\sqrt{\cosh \lambda}},
\end{equation}
with the Bohr-Sommerfeld phase $\phi_{BS}$ being the action of the 
map at the fixed point $z_0$, divided by $\hbar$.

\section{Semiclassical smoothing of the spectral decomposition of $\ket{z_0}$}
We rewrite equation (5) of reference \cite{Vergini_2017} for the 
case of an are-preserving map in the unit torus (we use the 
notation introduced in the letter)
\begin{equation}
    \sum |c_i|^2 g\left(\frac{2 (\phi-\tilde{\phi}_i)}{\lambda \beta}\right)\simeq K
    \tilde{F}(x) \left( 1 + \frac{1}{\lambda \sqrt{N}} 
  \sum \frac{\cos(\psi_j)}{\sqrt{A_j |L_j|}} \right),
\end{equation}
where the sum runs over the set of HOs with relevance $A_j< A_{max}$,
with $\beta=2/\ln(2 \pi N A_{max} )$ and $L_j$ the Lazutkin canonical
invariant. Moreover, $K=\sqrt{\pi/8} \beta/(1+\beta)$ and  
\begin{equation}
 g(y)=\frac{\sin(y)/y+\beta \cos(y)}{(1+\beta^2 y^2) (1+\beta)}
\quad {\rm for}\quad y\ne 0\quad {\rm and} \quad g(0)=1.
\end{equation}
\par
As $A_{max}$ increases, one has to include more HOs into the sum.
Then the fuction $g(2 (\phi-\tilde{\phi}_i)/\lambda \beta)$ is
closer to a delta function
and the eigenphases are better defined. 
\par
\begin{figure}[h] 
    \centering
    \includegraphics[width=0.65\linewidth]{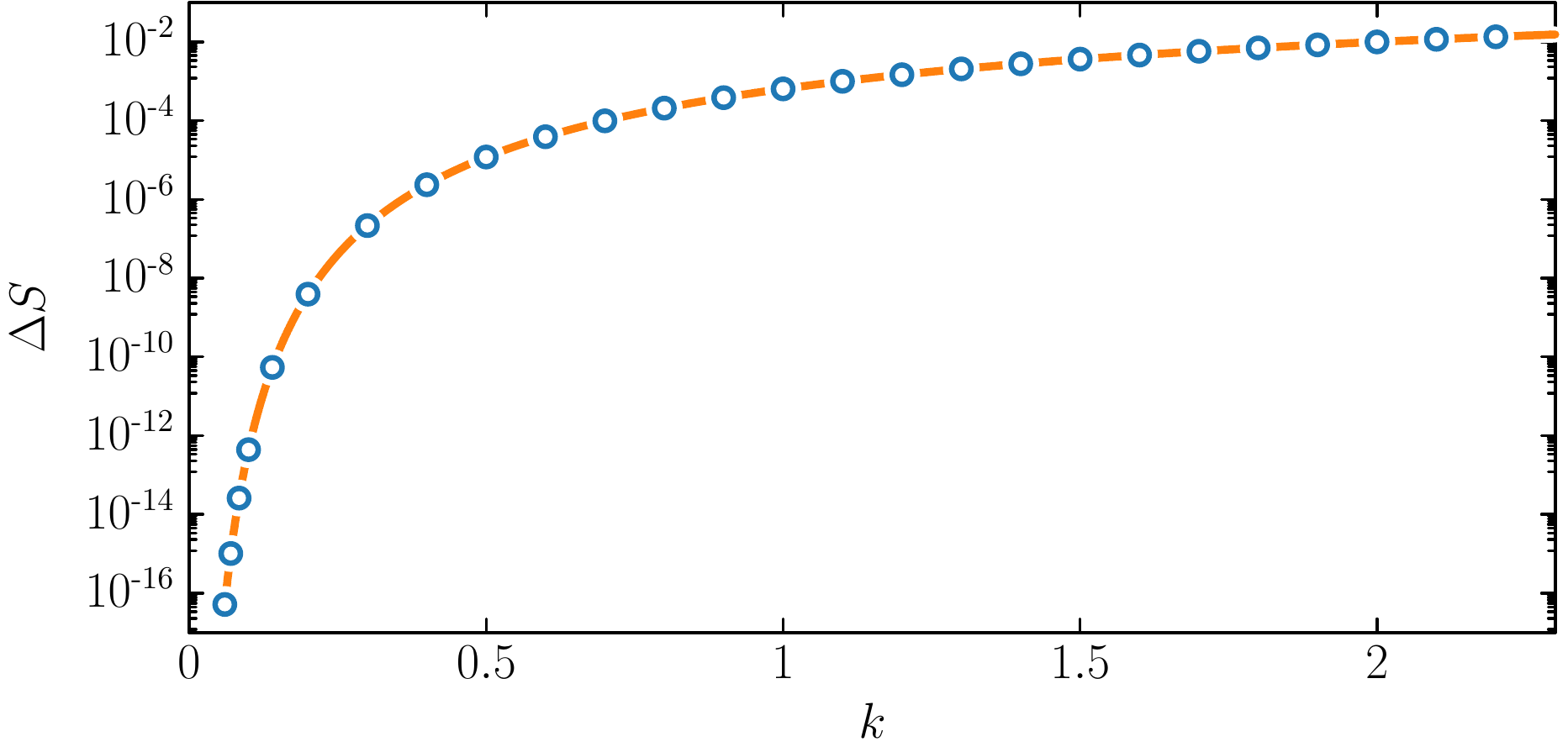} 
    \caption{The points are the calculated values of $\Delta S $ and the solid line corresponds to the estimated function of Eq.~(6) in the main text.
    \label{figDeltaS}}
\end{figure}
\section{Approximation of $\Delta S$ }
In Fig.~\ref{figDeltaS} we show the estimation Eq.~(6) (main text) of the canonical invariant $\Delta S=S_2-S_1$ (black shaded area in  Fig.~1(f)) as a function of $k$  
\section{The functions $\eta (x)$ and $\tilde{F}(x)$  }
\begin{figure}[h] 
    \centering
    \includegraphics[width=0.45\linewidth]{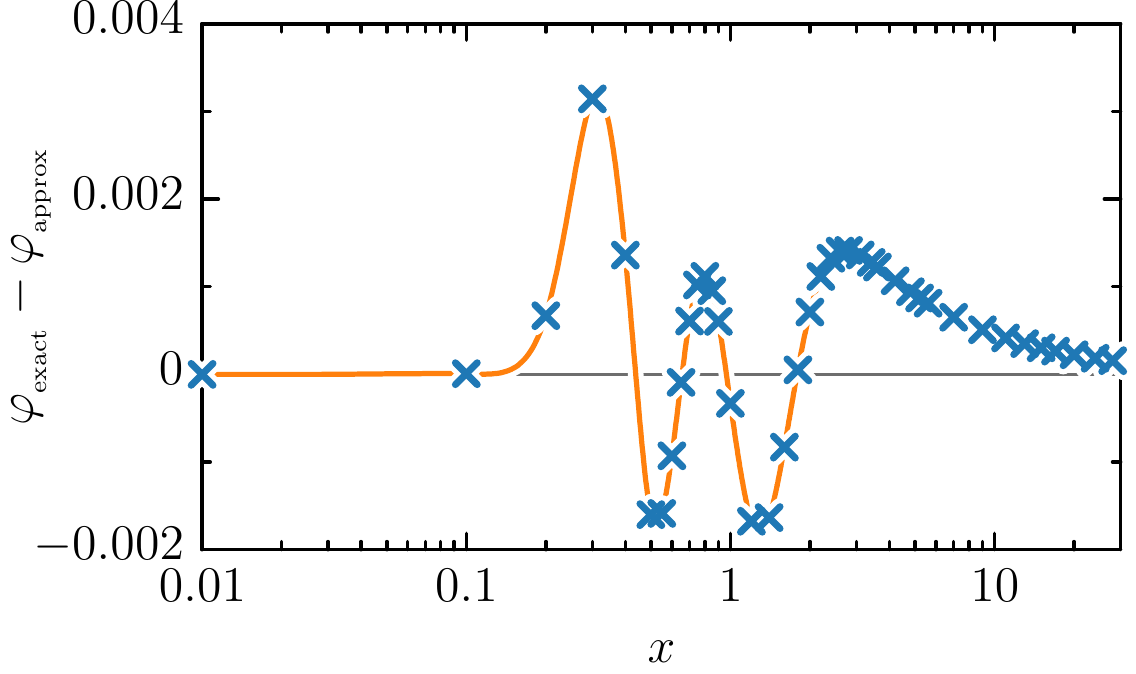} 
    \includegraphics[width=0.45\linewidth]{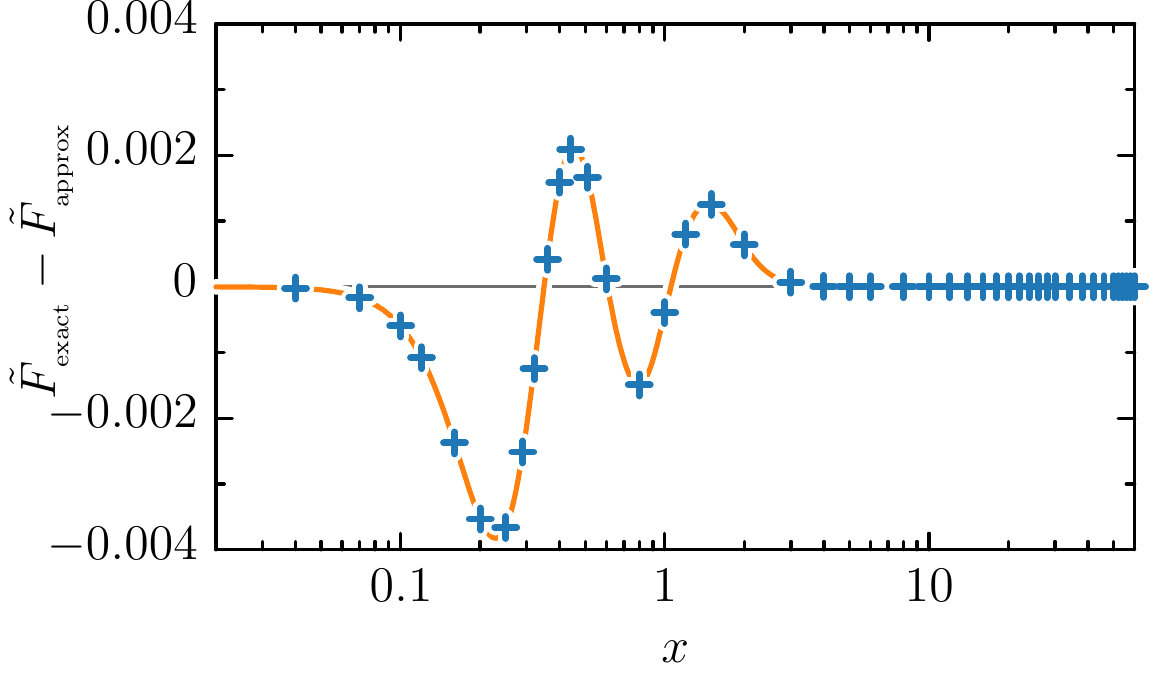} 
    \caption{Error in the approximation of $\varphi(x)$ and 
    $\tilde{F}(x)$.
    \label{varphiFourier}}
\end{figure}

The functions $\eta(x)$ and $\tilde{F}(x)$ are defined
in terms of the Fourier transform of
other functions \cite{Vergini_2017}. 
For this reason, in order to provide simple expressions here we  present accurate interpolation formulas.
\par
The function $\varphi(x)= x \eta(x)$ is the phase of the complex function
\begin{equation}
 \tilde{f}(x)= \frac{1}{\sqrt{\pi}}\int_{-\infty}^{\infty} {\rm e}^{-y/2} K_0({\rm e}^{-y}) {\rm e}^{{\rm i} x y} {\rm d}y,
 \end{equation}
with $K_0$ the modified Bessel function of zeroth order.
That is $ \tilde{f}(x)=  |\tilde{f}(x)| {\rm e}^{{\rm i} \varphi(x)}$,
where we have selected the branch of the phase satisfying $\varphi(0)=0$.
The function $\eta(x)$ verifies the following asymptotic behaviours
for $x \to 0$ and $|x| \to \infty$
\begin{equation}
   \eta(x)=\eta(0) -a x^2 + b x^4 + \order{x^6} \quad {\rm and} \quad
   \eta(x)= - \ln |x|+1+\frac{\pi}{4 |x|}+\order{1/x^2},
\end{equation}
with $\eta(0)\approx 3.53430\;63528$, $a\approx 5.38865\;58307$
and $b\approx 12.80436\;182$.
Later, we propose an interpolation formula satisfying 
the previous behaviours
\begin{equation}
    \eta(x)\approx -\ln \sqrt{x^2+1/2a}+1+A 
    \left(1+B x^4\right)^{-1/4}+ C 
    \left(1+z_1 x^6 \right)^{z_2 \ln(z_3+x^2)},
\end{equation}
with $z_1=423$, $z_2=-0.4337$ and $z_3=1.78$ free parameters fitted
to minimize the error of the approximation for $\varphi(x)$ in 
the intermediate region. 
In contrast, the constants $A$, $B$ and $C$ are related 
to $\eta(0)$, $a$ and $b$ as follows
\begin{equation}
    B=\left[ \frac{16}{\pi} (a^2-b) \right] ^{4/5}\approx 34.190,
\end{equation}
$A=\pi B^{1/4} /4 \approx 1.8992$, and $C=\eta(0)-\ln\sqrt{2a}-A-1\approx-0.5536$.
\par
The function $\tilde{F}(x)$ is defined by the integral equation
\begin{equation}
   \tilde{F}(x)= \sqrt{\frac{2}{\pi}} \int_0^{\infty} \frac{\cos(x y)}{\sqrt{\cosh(y)}} {\rm d}y,
\end{equation}
with asymptotic behaviours for $x \to 0$ and $|x| \to \infty$
\begin{equation}
 \tilde{F}(x)= \tilde{F}(0)- c x^2+\order{x^4} \quad {\rm and}  
 \quad  \tilde{F}(x)= \sqrt{\frac{2}{|x|}} {\rm e}^{-\pi |x| /2} 
 \left(1+\order{1/x^2}\right),
\end{equation}
with $\tilde{F}(0)\approx 2.09209\;92401$ and 
$c\approx 8.99462\;98154$. Later, we propose an interpolation 
formula satisfying the previous behaviors
\begin{equation}
 \tilde{F}(x)\approx  \frac{(D^2+x^2)^{-1/4}}{\sqrt{\cosh(\pi x)}}
 +E (1+z_4 x^4)^{- \ln \sqrt{z_5+x^2}},
\end{equation}
with $z_4=103$ and $z_5=1.8$ free parameters fitted
to minimize the error of the approximation for $\tilde{F}(x)$ 
in the intermediate region. 
In contrast, the constants $D\approx 0.31247$ and 
$E\approx 0.30316$ were derived from the relationships
$\tilde{F}(0)=E+1/\sqrt{D}$ and $c=(1+D^2\pi^2)/(4 D^{5/2})$.
\par
Figure~\ref{varphiFourier} shows the error of the previous
approximations.
\par
\section{Detailed phase space portrait for $k=1.447$}
\begin{figure}[h] 
    \centering
    \includegraphics[width=0.65\linewidth]{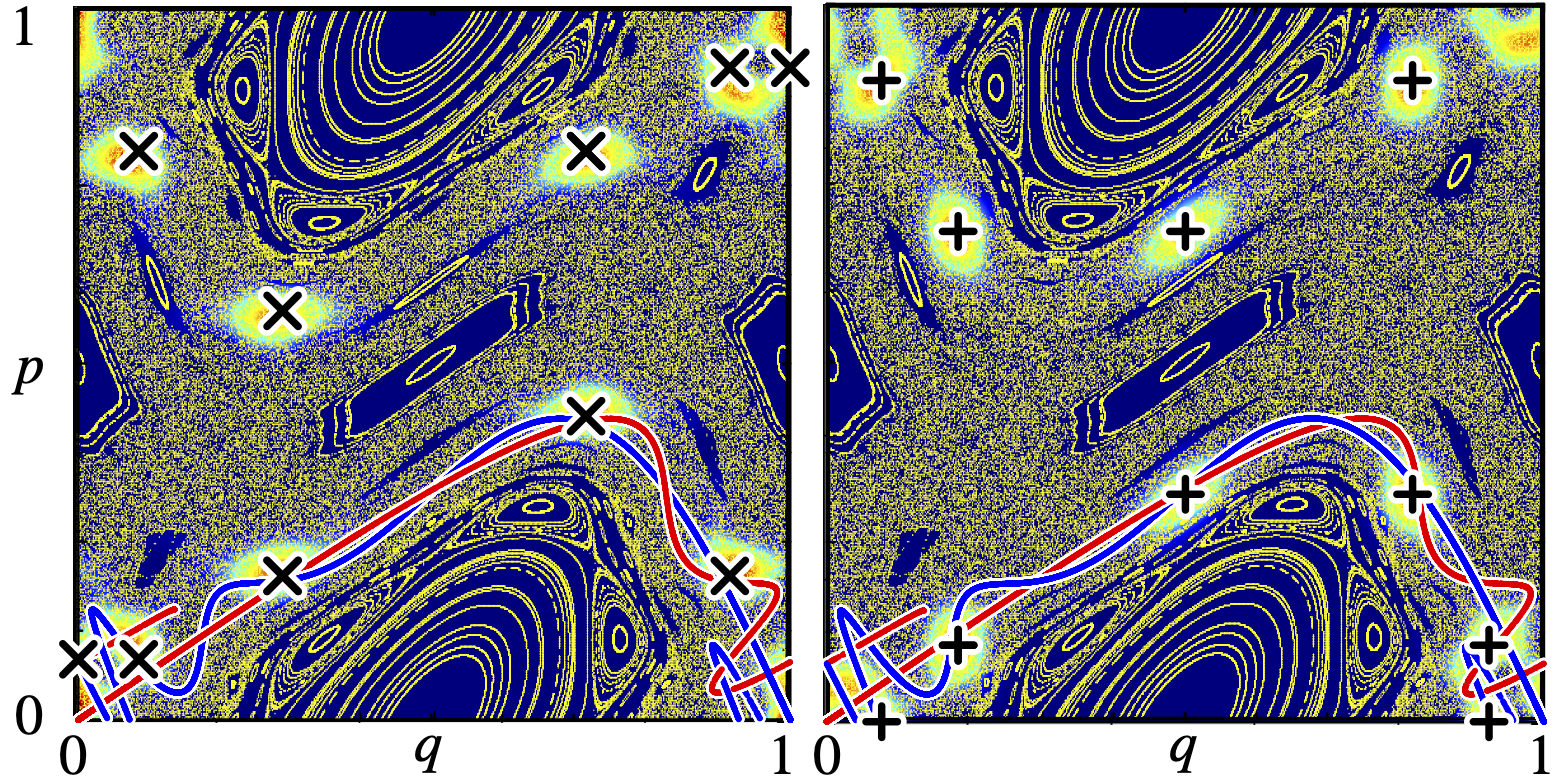}  
    \caption{Eigenfunctions No. 3 and 4. from Fig.~3 of the main text, with the corresponding satellite orbit in black symbols and the unstable (solid red) and stable (solid blue) manifolds. The yellow dots represent the classical phase space for $k=1.447$.
    \label{funcsupmat}}
\end{figure}

The infinite family of satellite POs of a given HO approximate
more and more the motion of the HO as the period of the PO goes
to infinity. This family consists of unstable POs with Bohr-Sommerfeld
phase well approximated by the condition $\psi_j(\phi)=2 \pi n$, with
$\psi_j(\phi)$ the pase of the given HO.
We have found that the shortest satellite PO of the first HO
generates a strong scar on the eigenstate number $3$ of 
Fig.~3 (of the main text), and the
shortest satellite PO of the second HO generates a strong scar
on the eigenstate number $4$. Fig.~\ref{funcsupmat} shows clearly 
these satellite POs, and their strong correlation with the 
Hussimis of states $3$ and $4$.

\section{Scaling of the participation ratio with the system size}
\begin{figure}[h] 
    \centering
    \includegraphics[width=0.85\linewidth]{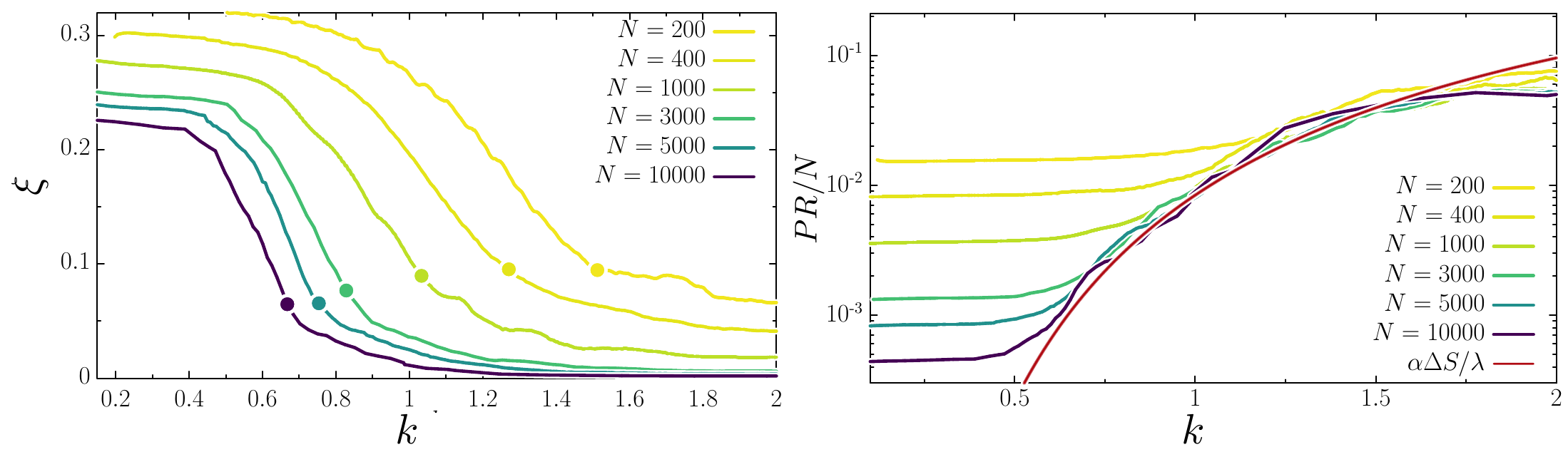}  
    \caption{(Left) Smoothed IPR $\xi$ as a function of $k$ for $N=200,400,1000,3000,5000, 10000$ (darkest shade is larger $N$). The dots correspond to the value of $k_{\rm break}$ obtained from $\Delta S/\hbar=3\pi/2$ [see main text].  (Right) $PR/N$ as a function of k for $N=200, 400,1000,3000,5000,10000$ (darker color means larger $N$). The red line represents $\alpha\Delta S/\lambda$ (we took $\alpha=14$). For all the curves a running average was performed to smooth out the fluctuations.
    \label{PR}}
\end{figure}

In Fig.~3 of the main text using a rescaling of the IPR $\xi$ we could show that a transition effectively takes place at an estimated value $k_{\text{break}}$. For clarity, and completeness in Fig.~\ref{PR} (left) we show a smoothing of the data obtained for $\xi$ (with no normalization) as a function of the (unrescaled) perturbation $k$, for several values of $N$. We observe that as $N$ grows, the step in the curves become steeper (and for smaller $k$). In fact, the -- correspondingly colored -- dots mark the $k_{\text {break}}$ value, where a clear displacement is observed, as it was shown in the inset of Fig.~3 [main text].

A complementary view of the IPR is provided by its inverse, i.e. the participation ratio or participation number (PR). It is usually associated with the area of Hilbert space occupied. If the system is fully chaotic then this saturates to a constant value proportional to $N$. 

In our case, the area filled by the chaotic layer is much smaller than unity (the area of the phase space) in 
the considered range of $k$. One should use an effective size $N_{\rm effective}$ instead of $N$, with
$N_{\rm effective}\approx N A_{\rm chaotic}$, and  where $A_{\rm chaotic}$ is the classical area of the chaotic layer which depends on $k$.
An elementary estimation, for small $k$, gives  
 $A_{\rm chaotic}\propto\, \Delta S  /\lambda$, where $\Delta S$ is the area of each
lobe (as represented in Fig.~1f), and $1/\lambda$ proportional to the number of
lobes along the broken separatrix. To clarify this point, in Fig.~\ref{PR} (right) we show
 $PR/N$ as a function of $k$ for a wide range of values  of $N$. It can be clearly observed that
for small $k$ values the scaling of $PR/N$ collapses onto the curve $\alpha \Delta S /\lambda$ (where $\alpha$ is an fitting prefactor). Furthermore, we observe that
the $PR$ of all the curves saturates at $N_{\rm effective}$ in accordance with our previous prediction.
On the other hand, in the small $k$ limit  $A_{\rm chaotic}< \xi_N$ (where $\xi_N$ is the normalization used in Fig.~3), and so  we have $PR/N\sim \xi_N/N$.
\section{Complementary data }
\begin{figure}[h] 
    \centering
    \includegraphics[width=0.65\linewidth]{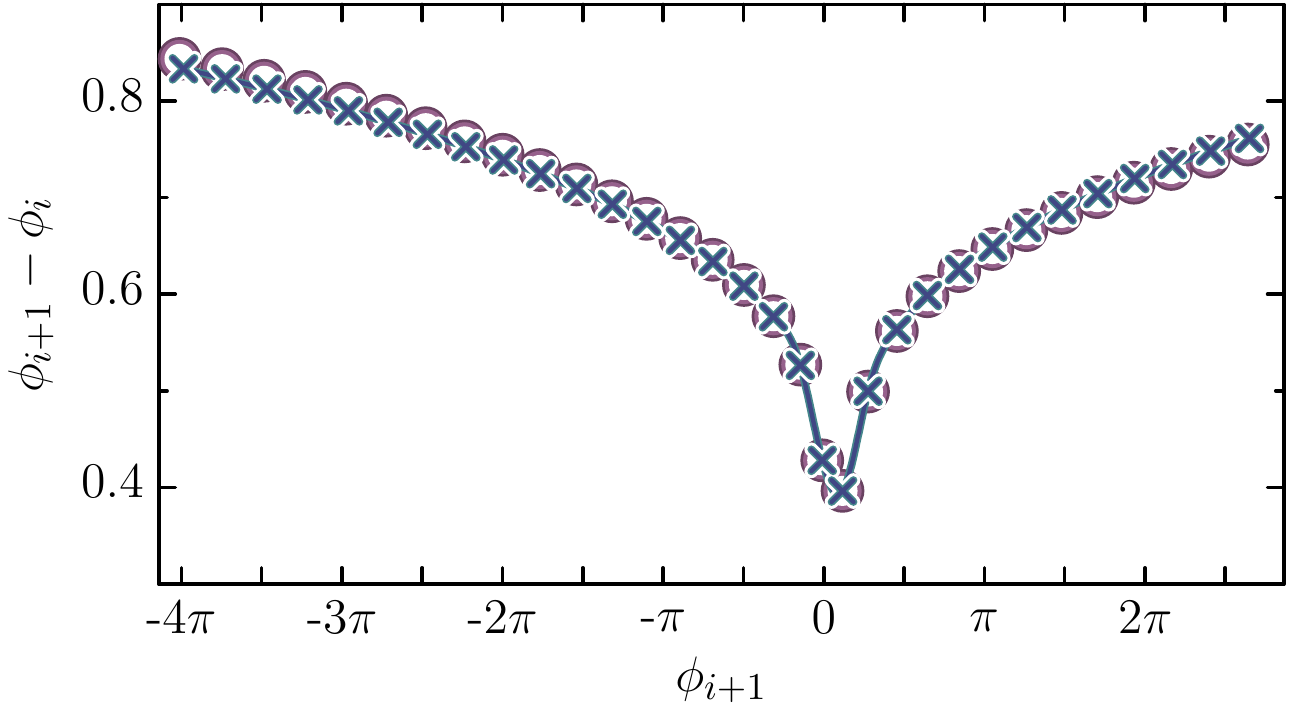}  
    \caption{ Level spacing singularity characteristic of ESQPT, for the standard map  with $N=1026$, $k=0.5$. Circles and crosses indicate quantum and 
    semiclassical calculation, respectively. The solid line is a cubic spline interpolation of the semiclassical calculation.
    \label{peine1026}}
\end{figure}

In this section we present data that supports the discussion presented in the main text. In Fig.~\ref{peine1026} we show the  logarithmic singularity in the level spacings  characteristic of the ESQPT, for a larger $N=1026$ value and the same perturbation $k=0.5$ as shown in Fig.~2 of the main text. The spectrum considered corresponds to the largest overlaps of the initial state $\ket{z_0}$ with the eigenstates of the map $U$. To compute the phase differences $\Delta\phi=\phi_{i+1}-\phi_i$, an unfolding procedure is necessary to account for contributions from different { Demkov structures  (see Fig.~2, main text)}. We note that  the semiclassical calculation  improves appreciably. This is expected because for $k<< k_{\text break}$ the error is of order $1/N$. For the perturbation value considered here we can estimate, by solving for $N=3/(4\Delta S)$ that the characteristic (singular) structure  will break at $N\approx 62900$.

\end{document}